\documentclass[aps,prb,superscriptaddress,showpacs,twocolumn,citeautoscript,notitlepage]{revtex4-1}

\usepackage{graphicx}
\usepackage{amsmath}
\usepackage{amssymb}
\usepackage[normalem]{ulem}
\usepackage{braket}
\usepackage[usenames,dvipsnames]{xcolor}

\newcommand{\bea}{\begin{eqnarray*}}
\newcommand{\eea}{\end{eqnarray*}}
\newcommand{\bne}{\begin{equation*}}
\newcommand{\ede}{\end{equation*}}

\newcommand{\bnen}{\begin{equation}}
\newcommand{\eden}{\end{equation}}
\newcommand{\bean}{\begin{eqnarray}}
\newcommand{\eean}{\end{eqnarray}}
\newcommand{\bsen}{\begin{subequations}}
\newcommand{\esen}{\end{subequations}}

\newcommand{\bna}{\begin{array}}
\newcommand{\eda}{\end{array}}
\newcommand{\bnm}{\begin{enumerate}}
\newcommand{\edm}{\end{enumerate}}
\newcommand{\bni}{\begin{itemize}}
\newcommand{\edi}{\end{itemize}}

\renewcommand{\vec}[1]{\text{\boldmath{$ #1 $}}}

\begin{document}

\title{
Impurity-assisted electric control of spin-valley qubits
in monolayer $\text{MoS}_2$}

\author{G. Sz\'echenyi}
\affiliation{Institute of Physics, E\"{o}tv\"{o}s University, 1518 Budapest, Hungary}

\author{L. Chirolli}
\email{luca.chirolli@imdea.org}
\affiliation{IMDEA Nanociencia, S-28049 Cantoblanco Madrid, Spain}

\author{A. P\'{a}lyi}
\affiliation{Department of Physics and 
MTA-BME Exotic Quantum Phases "Momentum" Research Group, 
Budapest University of Technology and Economics, 1111 Budapest, Hungary}

\date{\today}

\begin{abstract}

We theoretically study a single-electron spin-valley qubit in an electrostatically defined quantum dot in a  transition metal dichalcogenide monolayer, focusing on the example
of MoS$_2$. 
Coupling of the
qubit basis states for coherent control is challenging, as it requires a simultaneous flip of spin and valley.
Here, we show that a tilted magnetic field together with a
short-range impurity,
such as a vacancy, a substitutional defect, or an adatom, can give rise to a coupling between
the qubit basis states.
This mechanism renders the in-plane $g$-factor nonzero,
and allows to control the qubit with an in-plane ac electric field,
akin to electrically driven spin resonance. 
We evaluate the dependence of the in-plane $g$-factor
and the electrically induced qubit Rabi frequency on 
the type and position of the impurity.
We reveal highly unconventional features of the 
coupling mechanism, arising from symmetry-forbidden
intervalley scattering, in the case when the
impurity is located at a S site. 
Our results provide design guidelines for electrically controllable
qubits in two-dimensional semiconductors. 

\end{abstract}

\maketitle


\section{Introduction}

The electron spin in confined semiconductor quantum dots\cite{Hanson-rmp,Petta} (QDs) represents an ideal qubit system for 
encoding information at the quantum level, and a promising building block for quantum information processing.
All-electric manipulation of electron spins in QDs 
is enabled by spin-orbit interaction in two-dimensional 
electron systems as well as in nanowires
\cite{Flindt,Golovach,Rashba2008,Nowack-esr,Nadj-Perge2010,NadjPerge}.
The electronic valley degree of freedom\cite{Nebel}, relevant, e.g., in
silicon\cite{Maurand,Zwanenburg}, carbon nanostructures, and
transition metal dichalcogenides (TMDCs), has also been proposed
for quantum information processing purposes.
This degree of freedom could be utilized as a qubit on its own\cite{Culcer,Rohling,Laird}, 
or together with 
the electron spin, forming a combined spin-valley qubit\cite{Palyi-valley-resonance,Laird,Palyi-cnt-spinblockade,Li-edsr,FlensbergMarcus}.

The rise of two-dimensional (2D) materials and van der Waals heterostructures \cite{Novoselov,Roldan} promoted
2D semiconducting
TMDCs as alternative platforms for electronics, spintronics\cite{Zutic,Han}, and valleytronics\cite{Schaibley} 
applications, opening up new opportunities in nanoelectronics and optoelectronics with two-dimensional crystals\cite{Wang2012}. An appealing feature of TMDCs is the 
strong spin-orbit interaction, which is characteristic of both the valence and conduction bands\cite{Cappelluti13,Roldan14}, and arises due to the presence of the
heavy transition-metal atoms of the material. 
In particular, the broken inversion symmetry of monolayer (ML)
TMDCs gives rise to a strong spin-valley locking, whereby the Bloch
states close to the valleys have an out-of-plane spin polarization\cite{Kormanyos-mos2}. 
Furthermore, the possibility of electrostatically defining QDs in TMDCs such as MoS$_2$, WS$_2$ and WSe$_2$\cite{Song,Luo2017,Song2015,Javaid,KeWang,ZhuoZhiZhang} 
offers new opportunities
for spin-based quantum information processing\cite{Klinovaja,Kormanyos-mos2quantumdots,GuiBinLiu_intervalley,YueWu_spinvalley,Pearce_Burkard,BrooksBurkard}.
On the one hand, the nuclear-spin-free environment\cite{YueWu_spinvalley,Pearce_Burkard}, achievable via
isotopic purification, is expected to prolong the qubit lifetime compared to 
III-V materials such as GaAs -- 
in fact, this strategy has already proven to boost decoherence times
in diamond and silicon\cite{Itoh_isotope}.
On the other hand, the spin-orbit interaction present in TMDCs 
offers the possibility of efficient spin control via electric fields, which has potential
advantages over magnetic control.

In this work, we consider a single electron confined in a QD, which is 
electrostatically defined in the conduction band 
of a ML TMDC  (see Fig.~\ref{fig:setup}).
The two lowest-energy states in this setup form a spin-valley qubit:
the strong spin-orbit interaction and the broken inversion symmetry lock the spin to the valley
degree of freedom, and thereby the qubit basis states are characterized by
opposite spin in opposite valleys, $\ket{K \uparrow}$ and $\ket{K' \downarrow}$, in analogy with 
the spin-valley (or 'Kramers') qubit in carbon nanotubes\cite{FlensbergMarcus}.
An interesting feature of the spin-valley qubit is that it is difficult to induce
a coupling between the basis states, since that requires a simultaneous
flip of the spin and the valley.
On the one hand, this is an advantage, since it results in a suppressed
qubit relaxation, which might imply a prolonged qubit lifetime.
On the other hand, this makes it difficult to control the qubit with 
resonant excitation.

Here, we show that a  short-range impurity
(e.g., vacancy, substitutional atom, adatom)
in a ML-TMDC QD (see Fig.~\ref{fig:setup}) can couple
the basis states of the spin-valley qubit, and thereby allow
for resonant qubit control via an ac electric field, in the spirit
of electrically driven spin resonance.
The two main target quantities we calculate are 
(1) the  in-plane $g$-factor $g_{xx}$, which is made finite by the presence of 
a short-range impurity, and
(2) the Rabi frequency $\Omega_\text{R}$ 
characterizing the speed of the electrically induced
dynamics of the spin-valley qubit. 
We calculate how these two quantites depend on the system parameters, 
in particular their dependence 
$g_{xx}(x_0,y_0)$ and
$\Omega_\text{R}(x_0,y_0)$ on the relative position of the
impurity and the QD centre. 
Based on recent results\cite{GuiBinLiu_intervalley,Kaasbjerg2016,Kaasbjerg2017} 
on symmetry-forbidden intervalley scattering in ML-TMDCs, 
we reveal that the spatial patterns $g_{xx}(x_0,y_0)$ and
$\Omega_\text{Rabi}(x_0,y_0)$ depend drastically
on the type of the impurity (Fig.~\ref{fig:result}):
qualitatively different spatial patterns are obtained for a
defect located at a transition-metal site
(\emph{Mo-type impurity}, e.g., a Mo vacancy in MoS$_2$), 
and for one located at a chalcogen site
(\emph{S-type impurity}, e.g., an S vacancy in MoS$_2$). 
We discuss the role of symmetry-forbidden intervalley scattering
for adatoms as well (see Table \ref{tab:intervalley}).
We also highlight the fact that the impurity-induced coupling 
between the spin-valley qubit states increases as the spin-orbit
splitting decreases.
Therefore, we expect that the associated effects are most easily observable
in the ML-TMDC with the smallest 
conduction-band spin-orbit splitting, that is, in MoS$_2$.

The rest of the paper is organized as follows. In Sec.~\ref{setup} we introduce the setup and 
the ingredients of our model: the envelope function approximation, the spin-orbit and  Zeeman 
interactions, the impurity matrix elements, and the electrical drive. In Sec.~\ref{Sec:Intervalley} 
we describe intervalley scattering between the electronic states, and the important differences 
between the Mo-type and S-type impurities. In Sec.~\ref{Spin-valleyQubit} we introduce the 
spin-valley qubit. In Sec.~\ref{sec:gfactor} we derive the in-plane $g$-factor of the spin-valley 
qubit and in Sec.~\ref{Rabi} we describe  electrically driven Rabi oscillations of the qubit. 
In Sec.~\ref{sec:mos2} we quantify our generic results in the
special case of MoS$_2$, and conclude in Sec.~\ref{discussion}.

\section{Setup and model}
\label{setup}

\begin{figure}
\includegraphics[width=1\columnwidth]{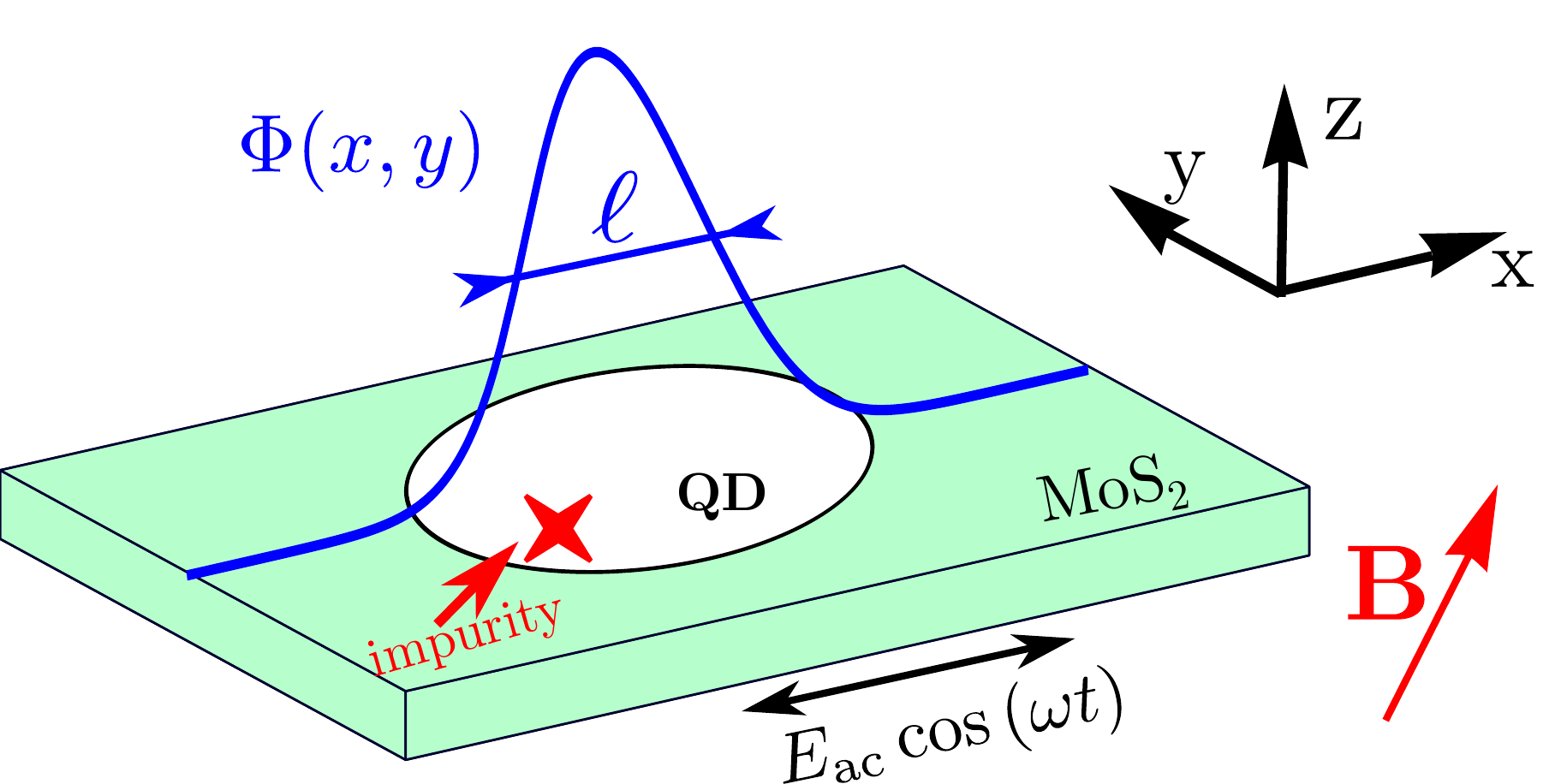}
\caption{\label{fig:setup}
\textbf{Electric control of a spin-valley qubit in a MoS$_2$ quantum dot.}
A circularly symmetric quantum dot holds a single electron, 
described by an envelope function $\Phi(x,y)$ with confinement length $\ell$. 
A short-range impurity is located in the dot, 
and a tilted  magnetic field $\vec B$ is applied. 
Electric control of the spin-valley qubit is realized with the in-plane ac electric 
field of amplitude $E_{\rm ac}$ and frequency $\omega$.
Control relies on the combination of the
electric excitation, the impurity, and the in-plane component of $\vec B$,
see Eq.~\eqref{eq:rabigeneric}.
}
\end{figure}

The setup we consider is shown in Fig.~\ref{fig:setup}. 
The material hosting the spin-valley qubit is a ML-TMDC,
e.g.,  MoS$_2$, lying in the $x$-$y$ plane. 
We consider an electrostatically defined QD, where a single electron
is confined with a parabolic, cylindrically symmetric
potential
\bean
\label{eq:vconf}
\mathcal{V}_\text{conf}(x,y)=\frac{1}{2}m^*\omega_0^2\left[(x-x_0)^2+(y-y_0)^2\right],
\eean
which is centered at the position $(x_0,y_0)$.
Here, $m^*$
is the effective mass of the electron, and  $\omega_0$ is the angular frequency
of the confinement. The typical length scale of the confinement is the oscillator length, 
defined as $\ell = \sqrt{\hbar / (m^*\omega_0)}$. In principle, the QD can be formed either
in the conduction band or in the valence band -- 
throughout the paper we use a terminology corresponding to the conduction band.
Here and henceforth, calligraphic font (e.g., $\mathcal V_\text{conf}$) is used to denote 
real-space Hamiltonians.

In the absence of spin-orbit interaction, magnetic fields, and impurites, 
the real-space single-electron Hamiltonian of this system reads
$\mathcal{H}_\text{QD} = \mathcal{K} + \mathcal{V}_\text{cr} + \mathcal{V}_\text{conf}$,
where the first two terms are the kinetic energy and the crystal potential,
respectively.
The confinement potential $\mathcal{V}_\text{conf}$ varies
slowly in space  ($\ell \gg a$, where $a$ is the lattice constant), therefore we can build the description of the electronic states upon 
the envelope-function Hamiltonian
\begin{equation} \label{envelope}
{\cal H}_\textrm{EF}=-\frac{\hbar^2}{2m^*}\left(\frac{\partial^2}{\partial x^2}+\frac{\partial^2}{\partial y^2}\right)+\mathcal{V}_\text{conf}(x,y).
\end{equation}
The eigenfunctions of ${\cal H}_\text{EF}$ have the form  $\Phi_{nm}(x,y)=\phi_n (x-x_0)\phi_m(y-y_0) $, where $\phi_{n} $ is the $n$th 1D harmonic-oscillator eigenstate, and the corresponding energy eigenvalue is $(n+m+1)\hbar\omega_0$.

Having the envelope functions $\Phi_{nm}$ at hand, we can express the corresponding
real-space wave functions as
\begin{equation} \label{nmvs_function}
\ket{nmvs}=\sum_\vec{k}c_{\vec k}^{nm}\ket{\Psi^{(1)}_{c,v\vec{K}+\vec{k}}} \chi_s.
\end{equation}
Here, 
the spin [valley] degree of freedom is incorporated via the spin quantum number
$s\in (\uparrow ,\downarrow)$
[valley index  $v \in (K,K') \equiv (+1,-1)$]
,
$c_{\vec k}^{nm}$ is the Fourier transform of  the envelope function 
$\Phi_{nm}(x,y)$,
and
$\chi_s$ is the spin wave function. 
Furthermore, 
$\ket{\Psi^{(1)}_{c,\vec{k}}}$ is the 
conduction-band 
Bloch function at the $\vec{k}$ point of the Brillouin zone, 
approximated up to first order in $k\cdot p$ theory
(see details in Appendices
\ref{sec:notation} and \ref{app:envelopefunction}).
Note that $c_{\vec k}^{nm}$  is a well localized function around $\vec{k}=0$, so, e.g., the 
state $\ket{nmKs}$  dominantly contains those 
Bloch functions whose wave 
vectors are within a small $\sim 1/\ell$ radius of the valley $K$. 
The  normalization condition is $\sum_{\vec k} | c_{\vec k}|^2 =1$. 
In what follows, we will 
 exploit time reversal symmetry to enforce
$\Psi^{(1)}_{c, \vec K'}(\vec r) = \left[\Psi_{c,\vec K}^{(1)}(\vec r)\right]^*$.

Beside the Hamiltonian
$\mathcal{H}_\text{QD}$, which incorporates
the kinetic energy, crystal potential and confinement potential discussed above,
we now add to our model the spin-orbit interaction, 
the magnetic field, the electric field, and the impurity.
To formulate this complete Hamiltonian, we use the basis defined in Eq.~\eqref{nmvs_function}.
In this basis, the real-space Hamiltonian $\mathcal{H}_\text{QD}$ is represented as
\begin{equation} \label{HamiltonianQD}
H_\textrm{QD}=\hbar\omega_0\left(a_x^{\dag}a_x+a_y^{\dag}a_y+1\right),
\end{equation}
with the creation operators defined via 
\begin{eqnarray}
a_x^{\dag}&=&\sum_{nmvs}\sqrt{n+1}\ket{(n+1)mvs}\bra{nmvs},\\
a_y^{\dag}&=&\sum_{nmvs}\sqrt{m+1}\ket{n(m+1)vs}\bra{nmvs}.
\end{eqnarray}

Even if the magnetic field is zero, the fourfold degenerate orbital states are split due to the 
intrinsic spin-orbit interaction, that plays a significant role in ML-TMDC. 
Typically, spin-orbit splitting $\Delta_{\textrm{SO}}$ is expected to 
exceed (or at least be comparable with) the orbital level spacing $\hbar\omega_0$
in a QD \cite{Kormanyos-mos2quantumdots}. 
We describe the spin-orbit splitting using the Hamiltonian 
\begin{equation} \label{so}
H_{\textrm{SO}}=-\frac{\Delta_{\textrm{SO}}}{2}\tau_3 s_z, 
\end{equation}
where 
 $\tau_3$ is the third Pauli matrix in valley space, defined as
$\tau_3 = \sum_{nmv}v\ket{nmv}\bra{nmv}$, and
$s_x$,$s_y$ and $s_z$ are spin Pauli matrices.
As a result of the spin-orbit interaction, the fourfold degeneracy of the orbital ground state is split to two 
Kramers doublets by an energy $\Delta_\textrm{SO}$, in a way that the ground-state Kramers doublet is  labelled with valley and 
spin indices  as $\ket{00K\uparrow}$ and $\ket{00K'\downarrow}$. 
The electron can occupy some superposition within the two-dimensional subspace of the lower-energy Kramers doublet,
i.e., the electron represents a Kramers qubit or spin-valley qubit.

The external magnetic field $\vec B = (B_\perp\cos{\varphi_B},B_\perp\sin{\varphi_B},B_z)$ is represented by the following Hamiltonian
\bean \label{magnetic}
H_B = \frac 1 2 \mu_B 
g_s \vec{B}\cdot\vec{s} +  \frac 1 2 \mu_B g_v B_z \tau_3
\eean
where the terms describe the spin and valley Zeeman splitting. Furthermore, $g_s\approx 2$ is the spin $g$-factor, $g_v$ is the material dependent valley $g$-factor, and $\mu_B$ is the Bohr-magneton. 
In the presence of $B_z$, the spin-valley qubit basis states
(i.e., the lower-energy Kramers doublet $\ket{00K\uparrow}$ and
$\ket{00K'\downarrow}$)  
are split by the energy $(g_v+g_s)\mu_B B_z$, so the corresponding out-of-plane $g$-factor $g_{zz}=g_v+g_s$ is finite and also material dependent. 
On the other hand, in an in-plane magnetic field, the spin-valley qubit remains 
degenerate, that is,
the in-plane $g$-factor ($g_{xx}=g_{yy}$) in a clean ML-TMDC QD is zero. 
The simple explanation behind this is that the 
Kramers-doublet have different spin and valley quantum numbers, 
and the in-plane $B$-field does not couple the different valleys.

Note that in Eq.~\eqref{magnetic}, 
we neglect  orbital effects of the magnetic field beyond the valley Zeeman 
effect, as those would not contribute to our results within the third-order perturbative
description we will apply below. 
Note, however, that the omitted orbital effects might be important in slightly different
settings, e.g., when the spin-valley physics of excited QD orbitals is described.

The impurity is modelled as a spin-independent scattering centre, described
by an electrostatic potential $\mathcal{U}_\text{imp}(\vec r)$.
We choose our reference frame such that the impurity 
is located at the origin; therefore, the relative position of the
impurity and the QD centre is represented by the 
QD centre location $\vec r_0 = (x_0,y_0)$ defined above. 
Because of the short-range character, the impurity couples 
different orbitals and valley effectively, hence it is expressed in 
our basis of Eq.~\eqref{nmvs_function}
as a dense matrix: 
\bean
H_\text{imp} = 
\sum_{nmv} \sum_{n'm'v'} \sum_s 
\tilde{\Delta}_{vv'}^{nmn'm'} 
\ket{nmvs} \bra{n'm'v's},
\eean
where we introduced the QD impurity matrix elements 
\bean
\label{eq:Delta}
\tilde{\Delta}_{vv'}^{nmn'm'} = \braket{nmvs | \mathcal{U}_\text{imp}|n'm'v's}.
\eean
In what follows, we will denote the absolute value and the
complex phase of the intervalley impurity matrix element as 
$\Delta_{KK'}^{nmn'm'} > 0 $ and 
$\varphi_{KK'}^{nmn'm'} \in (-\pi,\pi]$.


Finally, to control the spin-valley qubit, we apply an oscillating electric field  along the $x$ axis with amplitude $E_\textrm{ac}$ and driving angular frequency $\omega$, which is represented by the real-space Hamiltonian $\mathcal{H}_E = |{\rm e}|xE_\textrm{ac}\cos(\omega t)$.
In our model, we take this into account via 
\begin{equation}\label{electric}
H_E = |{\rm e}|\frac{\ell}{\sqrt{2}} (a_x+a_x^{\dag}) E_\textrm{ac}\cos(\omega t),
\end{equation}
where $\rm e$ is the electron charge.
This completes our model Hamiltonian
$H = H_\text{QD} + H_\text{SO} + H_B + H_\text{imp} + H_E$.

\section{Intervalley scattering}
\label{Sec:Intervalley}

Our goal is to describe the impurity-induced effects
(in-plane $g$-factor and electrically driven qubit Rabi oscillations)
using our QD model introduced above.
To proceed toward this goal, we need to relate the impurity-induced matrix elements
$\tilde{\Delta}_{vv'}^{nmn'm'}$ to the microscopic character of the impurity.
As we reveal below, Mo-type impurities and S-type impurities
imply qualitatively different intervalley matrix elements,
due to symmetry-forbidden intervalley scattering\cite{GuiBinLiu_intervalley,Kaasbjerg2016,Kaasbjerg2017}.

Combining Eqs.~\eqref{eq:Delta} and \eqref{nmvs_function}, 
we express the intervalley QD impurity matrix elements with
the bulk Bloch functions $\Psi_{\vec k}^{(1)}$ 
as follows:
\begin{eqnarray}\label{KK-general}
\tilde{\Delta}_{KK'}^{nmn'm'}=
\sum_{\vec{k},\vec{k'}} \left(c_{\vec k}^{nm}\right)^* c_{\vec k'}^{n'm'} M_{KK'}(\vec{k},\vec{k}'),
\end{eqnarray}
where we introduced the bulk intervalley impurity matrix elements
\bean \label{Mkk}
M_{KK'}(\vec{k},\vec{k}')=\braket{\Psi^{(1)}_{\vec{K}+\vec{k}}|\mathcal{U}_\textrm{imp}| \Psi^{(1)}_{\vec{K'}+\vec{k}'}}.
\eean
These bulk matrix elements and their dependence of 
the symmetry of the impurity have been characterized, and
also quantified for Mo and S vacancies in MoS$_2$, using 
density functional theory\cite{Kaasbjerg2016,Kaasbjerg2017}.
Note that it 
is experimentally established that such defects are dominant in 
ML-TMDC samples\cite{JinhuaHong, zhou, Lin2015,Vancso}.
The key findings of the symmetry analysis in 
Refs.~\onlinecite{Kaasbjerg2016,Kaasbjerg2017},
and their consequences for our problem,
are as follows. 

(i) \emph{Mo-type impurity.}
If the impurity 
is located at a transition-metal site, e.g., a Mo vacancy in MoS$_2$, 
then the bulk intervalley scattering impurity matrix element
connecting the $K$ and $K'$ points is finite in general, 
$M_{KK'}(\vec 0,\vec 0) \neq 0$.
In this case, since the Fourier transform $c_\vec k^{nm}$ of the
envelope function is localized around $\vec k = 0$, 
we obtain a finite result if we use 
the approximation
$M_{KK'}(\vec k,\vec k') \approx M_{KK'}(\vec 0,\vec 0)$
when evaluating Eq.~\eqref{KK-general}, 
and this will provide a good approximation.
This approximation results in 
a intervalley QD impurity matrix element
\begin{eqnarray}\label{KK-M}
\tilde{\Delta}_{KK'}^{nmn'm'}
&=&A\Phi^*_{nm}(0,0)\Phi_{n'm'}(0,0)M_{KK'}(\vec{0}),
\end{eqnarray}
where $A$ is the sample area.
Based on the numerical results of Ref.~\onlinecite{Kaasbjerg2016},
for a Mo vacancy in
MoS$_2$ 
we estimate $M_{KK'}(\vec 0,\vec 0) = 145 \, \text{eV} \mbox{\AA}^2/A$.

(ii) \emph{S-type impurity.}
If the impurity is located at a chalcogen site, e.g., an S vacancy
in MoS$_2$, then 
the bulk intervalley impurity matrix element, evaluated 
\emph{exactly} between $K$ and $K'$, vanishes:
$M_{KK'}(\vec 0,\vec 0) = 0$.
(Note that in the MoS$_2$ 
valence band, $M_{KK'}(\vec{0},\vec{0})$  is zero for both Mo and S vacancy.)
Numerical results in Ref.~\onlinecite{Kaasbjerg2016} also reveal that
$M_{KK'}(\vec k,\vec 0)$ is nonzero, and its absolute
value scales linearly with $k$ around $\vec k = 0$, that is, 
$|M_{KK'}(\vec k,\vec 0)| \propto k$.
Using symmetry arguments, we generalize this result 
in  Appendix \ref{app:symmetry}, where we show that
the bulk intervalley matrix element can be described for
short wave vectors as
\bean
\label{eq:sapprox}
M_{KK'}(\vec k,\vec k') \approx
\vec v \cdot (\vec k - \vec k'),
\eean
where
$\vec v = \gamma (1,-i) /A$
and $\gamma > 0$.
Based on the numerical results of Ref.~\onlinecite{Kaasbjerg2016},
for an S vacancy 
in MoS$_2$
we estimate $\gamma = 15 \, \text{eV} \mbox{\AA}^3$.
The approximation \eqref{eq:sapprox} is then used for
evaluating the intervalley
QD impurity matrix element in Eq.~\eqref{KK-general}, yielding
\begin{equation} \label{KK-X}
\tilde{\Delta}_{KK'}^{nmn'm'}
=\sum_{\vec{k},\vec{k'}} \left(c_{\vec k}^{nm}\right)^* c_{\vec k'}^{n'm'} \vec{v}\cdot\left(\vec{k}-\vec{k'}\right). 
\end{equation}

Due to the simple form of the harmonic-oscillator envelope functions, 
the intervalley QD impurity matrix elements \eqref{KK-M} and \eqref{KK-X}
can be evaluated analytically.
Note that here and henceforth we consider the case of a single impurity in the QD.
It is straightforward to generalize our results to an impurity ensemble and
provide a statistical description, see, e.g.,
corresponding work in the context of carbon nanotubes
\cite{Palyi_hyperfine,Palyi-valley-resonance, Szechenyi-maximalrabi}. 

Symmetry-forbidden intervalley
scattering, characteristic of S vacancies as described above, 
can also appear for other short-range impurities. 
A summary of our findings is shown in 
Table \ref{tab:intervalley}.
Naturally, for substitutional atoms replacing 
Mo [S] in the lattice, we expect to find a nonzero [zero]
direct intervalley impurity matrix element, as discussed in 
(i) [(ii)] above, as long as the substitutional atom 
does not change the corresponding symmetries.
Furthermore, we extend the symmetry analysis discussed in
this section and in Appendix \ref{app:symmetry} for 
the case of adatoms. 
The four typical high-symmetry locations for adatoms
are hollow-site (adatom on the out-of-plane axis
piercing the center of a Mo-S hexagon), 
bridge (adatom in the plane of the S-Mo-S bonds in a given unit cell), atop-Mo (adatom on the out-of-plane axis piercing a Mo atom)
and atop-S (adatom on the out-of-plane axis piercing a S atom).
In case of an atop-Mo [atop-S] adatom, intervalley scattering
is allowed [forbidden],
see  (i) [(ii)]. 
For a bridge adatom, we find that even though it does preserve
a symmetry, namely the S-Mo-S plane within the unit cell
remains a mirror plane, this does not give a restriction on the
intervalley matrix element. 
For a hollow-site adatom however, the three-fold rotational symmetry
around the axis, together with the symmetry of the conduction-band
wave function, implies that direct intervalley scattering is 
symmetry-forbidden, as in case (ii) above. 
 
\begin{table}
\caption{
Symmetry-forbidden 
 intervalley scattering by short-range impurities in MoS$_2$-type
materials.
`Forbidden' means that the bulk intervalley 
impurity matrix element vanishes,
$M_{KK'}({\bf 0},{\bf 0}) =0$ (see Eq. \eqref{Mkk}).  }
\begin{ruledtabular}
\begin{tabular}{l | c}
impurity type & direct intervalley scattering \\\hline
Mo vacancy & allowed \\
S vacancy & forbidden \\
Mo-type substitutional & allowed \\
S-type substitutional & forbidden \\
atop-Mo adatom & allowed \\
atop-S adatom & forbidden \\
hollow-site adatom & forbidden \\
bridge adatom & allowed 
\end{tabular}
\end{ruledtabular}
\label{tab:intervalley}
\end{table}

\section{Spin-valley qubit}
\label{Spin-valleyQubit}

We now have all the elements to characterize a spin-valley qubit in a ML-TMDC quantum dot. 
We consider the perturbative case,
when the basis states of the spin-valley qubit are 
energetically well separated from the other states, 
i.e., the coupling matrix elements between the 
qubit and the other states are much smaller 
than their energy difference:
\begin{equation}\label{hierarchy}
\Delta_{\textrm{SO}}, \hbar\omega_0 \gg\begin{cases}
    ~g_s\mu_BB_x \\
    ~g_v\mu_BB_z \\
    ~\Delta_{KK'}^{nmn'm'}\\
    ~|{\rm e}| E_\textrm{ac} \ell.
      \end{cases}
\end{equation}

Therefore we treat $H_0 = H_\text{QD} + H_{\textrm{SO}}$
as the unperturbed Hamiltonian and 
$H_1 = H_B+H_\text{imp}+H_{E}$ 
as the perturbation. 
The two lowest-energy eigenstate of 
the unperturbed Hamiltonian 
$H_0$  are
$\ket{00K\uparrow}$ and $\ket{00K' \downarrow}$;
we call these the \emph{unperturbed spin-valley qubit basis states}.
Below, we consider the undriven case $H_E = 0$ to
derive the in-plane $g$-factor, and
the driven case with finite $H_E$ to describe
the electrically induced qubit Rabi oscillations. 
In both cases, we apply Schrieffer-Wolff perturbation 
theory\cite{Winkler,Romhanyi} 
to derive a $2\times 2$ effective Hamiltonian 
for the spin-valley qubit. 
In both cases, the perturbed qubit basis state
associated to $\ket{00K\uparrow}$ and
$\ket{00K'\downarrow}$ will be denoted
by $\ket{\Uparrow}$ and $\ket{\Downarrow}$, respectively.


\begin{figure}
\includegraphics[width=1\columnwidth]{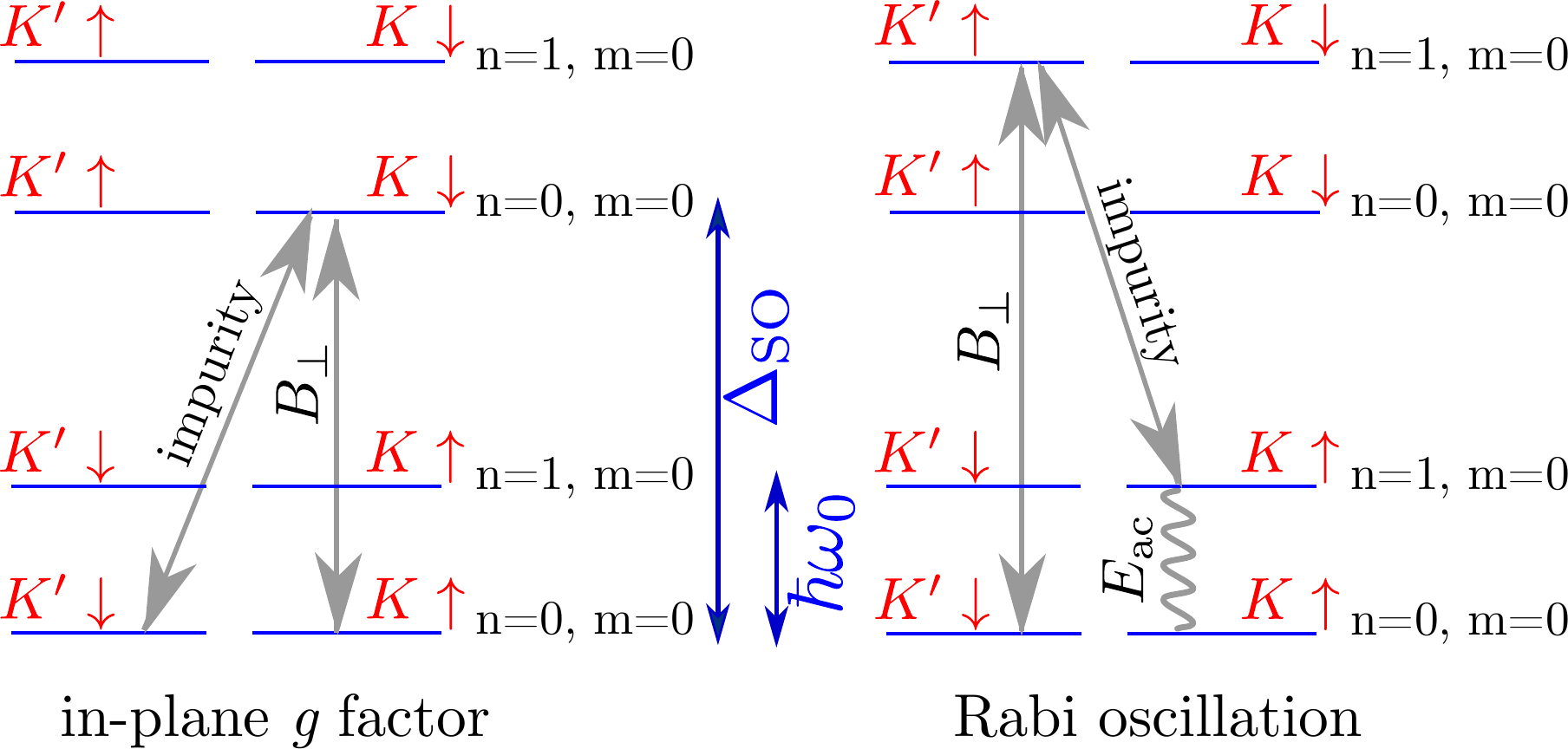}
\caption{\label{fig:PTlevels}
{\bf In-plane $g$-factor and electrically induced Rabi 
oscillations of a spin-valley qubit.}
Blue horizontal lines show the energy spectrum of the
unperturbed quantum dot, set by 
orbital level spacing $\hbar \omega_0$
and spin-orbit splitting $\Delta_\text{SO}$.
Lowermost two blue lines are  basis states of the
spin-valley qubit.
Gray arrows are perturbation matrix elements
(impurity, in-plane magnetic field, ac electric field)
that induce second-order static (a) or 
third-order dynamic (b) mixing of the 
qubit basis states. 
This mixing leads to a finite in-plane $g$-factor (a) and
electrically driven Rabi oscillations (b) of the qubit.}
\end{figure}

\section{In-plane $g$-factor}
\label{sec:gfactor}

In a clean dot, the in-plane magnetic field 
does not couple the basis states of the
spin-valley qubit, so the in-plane $g$-factor $g_{xx}$ is zero.
However, the impurity-induced intervalley matrix element
combined with a finite in-plane magnetic field
does couple the qubit basis states via intermediate states
at higher energies, see Fig.~\ref{fig:PTlevels}a.
Therefore, in the presence of impurities, 
a finite in-plane $g$-factor is expected for the spin-valley qubit.

To evaluate the in-plane $g$-factor, we use second-order
Schrieffer-Wolff perturbation theory\cite{Winkler}.
From that, we obtain the effective qubit Hamiltonian
\bean
H_\text{q} = H_\text{q}^{(1)} + H_\text{q}^{(2)},
\label{eq:h2bparallel}
\eean
where
\begin{eqnarray}
\label{eq:hq1}
H^{(1)}_\textrm{q}  
&=&  
\frac{1}{2}\mu_B(g_s+g_v)B_z\sigma_z
\\
H^{(2)}_\textrm{q}
&=&
-\frac{\mu_B g_s B_\perp \Delta^{0000}_{KK'}}{\Delta_{SO}}
\sigma(\varphi^{0000}_{KK'}+\varphi_B),
\label{eq:hq2}
\end{eqnarray}
$\sigma(\varphi)=\sigma_x \cos \varphi - \sigma_y \sin \varphi$,
and $\sigma_x$, $\sigma_y$, $\sigma_z$ are Pauli matrices acting on 
the perturbed qubit basis states
$\ket{\Uparrow}$ and $\ket{\Downarrow}$. 
The second-order term
 $H^{(2)}_\text{q}$ in Eq.~\eqref{eq:hq2} is interpreted
as a Zeeman interaction of the qubit with the in-plane magnetic field
$B_\perp$, 
which is turned on by the presence of the impurity. 
Fig.~\ref{fig:PTlevels}a illustrates one term
in the perturbative sum that generates
$H^{(2)}_\text{q}$, i.e., one path contributing to this
interaction: the lowest two blue lines represent
the unperturbed qubit basis states, other blue lines represent
other eigenstates of the unperturbed Hamiltonian, 
whereas
the gray arrows represent perturbation matrix elements 
connecting those states.

From Eq.~\eqref{eq:h2bparallel}, 
we see that the presence of the impurity does not 
affect the out-of-plane $g$-factor $g_{zz}$, at least in this 
order of perturbation theory. 
Furthermore, the 
in-plane $g$-factor is expressed as 
\bean\label{Gxx}
g_{xx} = 
2 g_s 
\frac{\Delta_{KK'}^{0000}}{\Delta_{SO}}.
\eean
Note that this is essentially the same result as obtained by 
Flensberg and Marcus\cite{FlensbergMarcus} for carbon nanotubes, 
see their Eq.~(5).

In the rest of this section, we
consider impurities that preserve the threefold rotational
symmetry of the lattice, either with a rotation axis 
containing an Mo atom (e.g., a Mo vacancy), 
or with a rotation axis containing
an S atom (e.g., an S vacancy). 
We 
 show that the dependence of the in-plane $g$-factor 
 $g_{xx}$
on the location $\vec r_0$ of the QD center with
respect to the vacancy is
qualitatively different for the Mo and S types.

The in-plane $g$-factor is 
governed by the intervalley impurity matrix element 
$\Delta_{KK'}^{0000}$.
For a Mo-type impurity, 
Eq.~\eqref{KK-M}
implies that this matrix element
 inherits the spatial dependence 
of the squared wave function at the impurity position:
\begin{equation} \label{KK_MC}
{\Delta_{KK'}^{0000}}=\frac{A}{\ell^2\pi}e^{-\frac{r_0^2}{\ell^2}}M_{KK'}(\mathbf{0},{\bf 0}).
\end{equation}
Due to Eq.~\eqref{Gxx}, $g_{xx}$ inherits
the same Gaussian spatial dependence:
\bean
\label{eq:gxxmo}
g^{\text{Mo}}_{xx} = g_{xx,\text{max}}^\text{Mo}
e^{-\frac{r_0^2}{\ell^2}},
\eean
with 
\begin{equation}\label{gmaxM}
g_{xx, {\rm max}}^\text{Mo}=\frac{2g_sAM_{KK'}(\mathbf{0},{\bf 0})}{\ell^2\pi\Delta_\textrm{SO}}.
\end{equation}

To evaluate $\Delta_{KK'}^{0000}$ for 
the case of an S-type impurity, we use Eq.~\eqref{KK-X}, 
and there we need the Fourier components of the 
ground-state QD orbital:
\bean
\label{ccoe}
c_{\vec{k}}^{00}=\frac{2\sqrt{\pi}\ell}{\sqrt{A}}e^{-\frac{\ell^2|\vec k|^2}{2}}e^{-i \vec{r}_0 \cdot \vec k}.
\eean
(Recall that our reference frame is chosen such that
the impurity is located at the origin and the QD center is displaced
to $\vec r_0$.)
Then, the intervalley matrix element for the S-type impurity
reads
\begin{equation}\label{KK_XC}
{\Delta_{KK'}^{0000}}=
\frac{2}{\pi}
\frac{\gamma}{\ell^4}r_0e^{-\frac{r_0^2}{\ell^2}},
\end{equation}
which implies via Eq.~\eqref{Gxx} that
\bean
g_{xx}^\text{S} = g_{xx,\text{max}}^\text{S}
\sqrt{ 2 e } \frac{r_0}{\ell} e^{-\frac{r_0^2}{\ell^2}},
\eean
with
\begin{equation}\label{gmaxX}
g_{xx,{\rm max}}^\text{S}=
\frac{2\sqrt{2}}{\pi \sqrt{e}}
\frac{g_s\gamma}{\ell^3\Delta_\textrm{SO}}.
\end{equation}

\begin{figure}
\includegraphics[width=0.9\columnwidth]{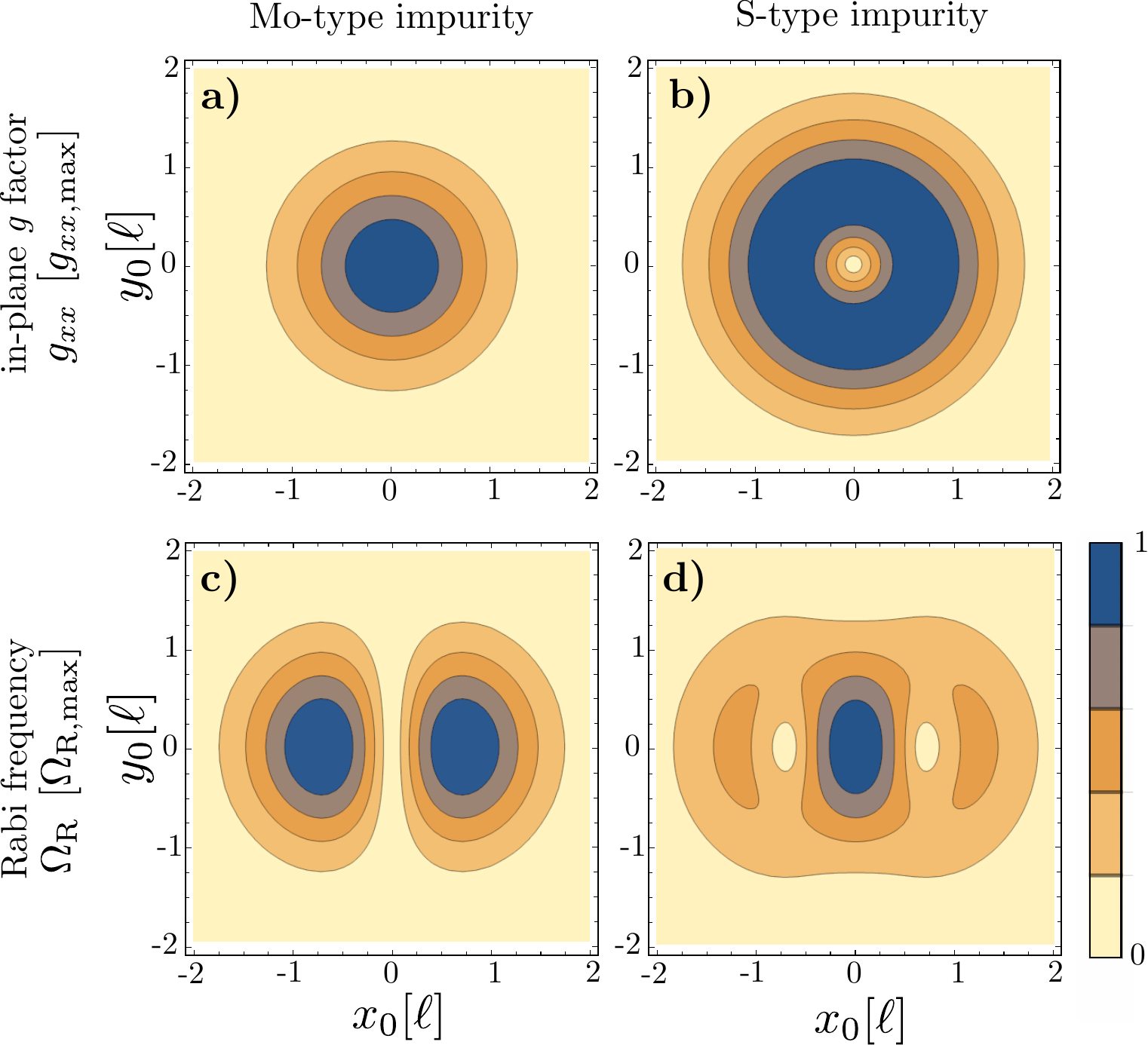}
\caption{\label{fig:result}
(Color online)
{\bf Dependence of in-plane $g$-factor and electrically driven
qubit Rabi frequency on impurity type and position.}
Upper row: in-plane $g$-factor for
a Mo-type (a) and a S-type (b) impurity, as functions of the
relative position of the impurity and the quantum-dot center. 
Bottom row: electrically driven qubit Rabi frequency 
for a Mo-type (c) and a S-type (d) impurity, 
for a driving electric field along the x axis. 
Color-code units for (a), (b), (c), (d) are
defined in Eqs.~\eqref{gmaxM}, \eqref{gmaxX}, \eqref{rabimaxM},
and \eqref{rabimaxX}, respectively.  
}
\end{figure}

Clearly, the dependencies of $g_{xx}^\text{Mo}$ and
$g_{xx}^\text{S}$ 
on $r_0$ are qualitatively different, which is
illustrated in Fig.~\ref{fig:result}a and b. 
Both plots show circular symmetry: the results
depend only on the length of $\vec r_0$ but 
not on its direction, see Eq. \eqref{KK_MC} and \eqref{KK_XC}. 
In case of the Mo-type impurity, Fig.~\ref{fig:result}a,
the in-plane $g$-factor decreases monotonically  
as $r_0$ is increased, following the 
Gaussian probability density of the envelope function at the position of
the impurity.
However, the in-plane $g$ factor due to an S-type impurity,
Fig.~\ref{fig:result}b, 
is different, it is zero if the impurity is at the QD center,
and reaches its maximal value $g_{xx,\text{max}}^\text{S}$
if the impurity is at a distance
$r_0 =  \ell/\sqrt{2}$ from the QD center. 

These results imply that
in principle, the in-plane $g$-factor in a MoS$_2$ QD with a 
short-range impurity
can be tuned electrically, by reshaping or 
replacing the QD as the voltages on the confinement
gates are changed. 
For example, in the presence of an S-type impurity, the
impurity-induced intervalley coupling, 
and thereby the corresponding decoherence processes,
could be suppressed by electrostatically 
tuning the QD to $(x_0,y_0) = 0$, 
i.e., without spatially separating the impurity
and the electron.
Also, the dependence of the in-plane $g$ factor
on the impurity position is qualitatively different 
for Mo-type and S-type impurities.
This implies that experiments, where, e.g., magneto-transport
spectroscopy maps the $g$ factor as a function of
QD shape and location, could reveal information
about the type, number, and position of the impurities
in the QD.

\section{Electrically driven qubit Rabi oscillations}
\label{Rabi}

Here we show that the spin-valley qubit defined in the 
MoS$_2$ QD can be coherently controlled by 
an ac electric field, if a tilted magnetic field is applied, and
at least one short-range impurity is present in the dot.
In this case, the ac electric field induces qubit Rabi oscillations,
when the qubit evolves coherently and cyclically 
between the two basis states $\ket{\Uparrow}$ and 
$\ket{\Downarrow}$,
if the driving frequency $\omega$ is chosen to be resonant with the
qubit's Larmor frequency $ \omega_q $. 
Our goal is to calculate the Rabi frequency characterizing
these oscillations. 

Similarly to the preceding section, 
here we also map our
Hamiltonian $H = H_0 + H_1$,
acting in an infinite-dimensional Hilbert space, 
to a two-dimensional effective qubit Hamiltonian.
Note that in this case, $H_1$ is time dependent 
due to the presence of the ac electric field in $H_E(t)$. 
We use time-dependent Schrieffer-Wolff perturbation 
theory\cite{Romhanyi},
treating $H_1$ as the perturbation,
to obtain the qubit Hamiltonian. 
We find that the 
leading-order time-dependent term in the effective qubit
Hamiltonian appears in the 3rd order of perturbation theory,
and reads
\begin{equation}
H_\textrm{q}^{(3)}(t) =
\hbar \Omega_\text{R} \cos(\omega t)
\sigma(\varphi^{0010}_{KK'}+\varphi_B),
\label{eq:hqubit3}
\end{equation}
where
\begin{equation}
\label{eq:rabigeneric}
\hbar \Omega_\text{R} = 
\sqrt{2}
\frac{(|{\rm e}| E_{ac} \ell) (g_s \mu_B B_\perp) \Delta^{0010}_{KK'} }{\Delta_{SO} \hbar \omega_0}.
\end{equation}
This coupling appears due to the interplay of the in-plane 
magnetic field, 
the ac electric field and the impurity.
Fig.~\ref{fig:PTlevels}b illustrates one of the three-step paths
via virtual intermediate states that contribute to this
coupling. 
Note that the fact that the only intervalley impurity 
matrix element appearing in Eq.~\eqref{eq:hqubit3}
is $\Delta_{KK'}^{0010}$ is due to the choice
that the driving electric field is along the $x$ direction and that
the confinement potential is parabolic. 

We consider the magnetic-field hierarchy 
$B_z\gg B_\perp \frac{\Delta^{0000}_{KK'}}{\Delta_{SO}}$, 
for which the qubit Larmor frequency 
is dominated by the out-of-plane component
of the magnetic field. 
In this case, the qubit Larmor frequency
is given by $\omega_q \approx \mu_B(g_s+g_v)B_z$. 
We want to describe the electrically induced dynamics of this qubit with 
the initial state being $\ket{\Downarrow}$. 
Upon resonant or almost-resonant driving, 
$\omega \approx \omega_q$, and assuming that the driving is weak, 
the dynamics remains in the two-dimensional space of the spin-valley qubit, 
and the qubit will show simple Rabi oscillations, 
characterized by the Rabi frequency $\Omega_\text{R}$ 
given by Eq.~\eqref{eq:rabigeneric}. 

From Eqs.~\eqref{eq:rabigeneric},
the Rabi frequency is proportional to the impurity matrix element 
$\Delta^{0010}_{KK'}$, which can be calculated for a Mo-type
impurity using Eq. (\ref{KK-M}). 
We find that the dependence of the 
Rabi frequency as a function of the impurity position 
is given by
\bean
\label{eq:rabimo}
\Omega_\textrm{R}^{\rm Mo} = \Omega_\textrm{R,max}^\textrm{Mo}  \sqrt{\frac{e}{2}}\frac{|x_0|}{\ell} e^{-\frac{r_0^2}{\ell^2}},
\eean
where the maximal value of the Rabi frequency is
\bean \label{rabimaxM}
\Omega_\textrm{R,max}^\textrm{Mo} = \sqrt\frac{2}{e}\frac{|e| E_\text{ac}g_s \mu_B B_\perp}{\pi\ell\Delta_{SO} \hbar ^2\omega_0} AM_{KK'}(\mathbf{0},\mathbf{0}).
\eean

The dependence of the Rabi frequency Eq.~\eqref{eq:rabimo}
on the impurity position 
is shown in Fig.~\ref{fig:result}c.
The key features are: 
(i) The Rabi frequency is maximized when 
the position of the impurity is $\vec r_0 =(\pm\frac{\ell}{\sqrt{2}},0)$.
(ii) The Rabi frequency decreases when the impurity
is moved outside from the QD. 
(iii) The Rabi frequency is exactly zero if the impurity is placed 
in the QD center, or on the line $x_0=0$. 

Feature (iii) can be explained as follows. 
The external ac electric field along the $x$ axis induces a spatial oscillation of the electron in the $x$ direction, therefore the relative position of the impurity and the QD also oscillates with a small 
$\ll \ell$ amplitude. 
Since, as seen in Fig.~\ref{fig:result}a,  the in-plane $g$ factor
 $g_{xx}$ depends on the 
relative position according to 
the spatial oscillation of the wave function 
leads to a temporal oscillation of $g_{xx}$, 
driving a Rabi oscillation between
spin-valley qubit basis states. 
However, in the vicinity of the  line $x_0=0$,
$g_{xx}$ depends on $x_0$ quadratically, therefore 
the ac electric field does not generate a time-dependent
$g_{xx}$ in first order, and therefore the Rabi frequency is zero. 

If the impurity in the QD is S-type, then the Rabi frequency is 
expressed from 
Eqs.~\eqref{eq:rabigeneric},
\eqref{KK-X}, \eqref{ccoe}, and 
\bean
c_{\vec k}^{10} = \frac{8\sqrt{\pi}i\ell^2}{\sqrt{A}}k_x e^{-\frac{\ell^2|\vec k|^2}{2}}e^{-i \vec{r}_0 \cdot \vec k}.
\eean
as
\begin{eqnarray}
\Omega_\textrm{R}^{\rm S} &=& \Omega_\textrm{R,max}^\textrm{S} |1-2x_0(x_0+iy_0)/\ell^2| e^{-\frac{r_0^2}{\ell^2}},
\end{eqnarray}
where 
\bean \label{rabimaxX}
\Omega_\textrm{R,max}^\textrm{S} = \frac{2|e| E_{ac}g_s \mu_B B_\perp}{\pi\ell^2\Delta_{SO} \hbar^2 \omega_0} \gamma,
\eean
The spatial dependence of the Rabi frequency
is plotted in Fig.~\ref{fig:result}d;
the main features are as follows.
(i) the Rabi frequency reaches its maximal value 
if the impurity is in the centre of the QD.
(ii) There is no coherent transition if the impurity 
is placed at the position 
$\vec r_0 =(\pm\frac{\ell}{\sqrt{2}},0)$.
(iii) The Rabi frequency has a local maximum if the monovacancy is at the position $\vec r_0 =(\pm\sqrt{\frac 3 2 }\ell,0)$. 
We note that the simple explanation of the previous paragraph,
applied for the Mo-type impurity and 
based on the adiabatic modulation of the in-plane
$g$-factor, does not explain features (i) and (iii) 
in the case of an S-type impurity.
We anticipate that in this case, the relation between
the $g$-factor and the qubit dynamics
can be characterized by a combination of 
the `Zeeman-modulation' and `iso-Zeeman-modulation' 
mechanisms\cite{Crippa}.

\section{Quantifying the results for MoS$_2$}
\label{sec:mos2}

In this work, the in-plane $g$-factor and 
the electrically driven Rabi oscillations are described
in the perturbative regime, see Eq.~\eqref{hierarchy},
where the spin-orbit splitting 
$\Delta_\text{SO}$ is a large energy scale.
In this regime, the coupling between the spin-valley qubit
basis states decreases as the spin-orbit splitting is increased. 
Among the semiconductor ML-TMD materials, 
MoS$_2$ has the smallest spin-orbit 
splitting\cite{Kormanyos-mos2quantumdots}.
Therefore, this seems to be the material best suited to 
observe the effects predicted here. 
Here we quantify the in-plane $g$-factor and Rabi frequency 
for a single-electron spin-valley qubit in a ML MoS$_2$ QD.

We consider a QD with orbital level spacing 
$\hbar \omega_0 \approx 0.5$ meV.
Using $m^* \approx 0.5~m_\textrm{e}$, 
the oscillator length is $\ell \approx 17.5$~nm. 
According to numerical studies of ML MoS$_2$,
the spin-orbit gap is $\Delta_\textrm{SO}=3$ meV and the 
valley $g$ factor is 
$g_v\approx 4$\cite{Kormanyos-mos2quantumdots}. 
We assume a magnetic field with in-plane component $B_\perp=1$ T,
and a driving electric field with amplitude $E_\text{ac}=10$ kV/m. 
The latter value implies that the amplitude of the electron’s spatial 
oscillations is 6 nm, smaller than the oscillator length $\ell$, 
so the perturbative treatment described above is 
justified.

If the Mo-type impurity is a Mo vacancy, 
and it is placed at the QD center, then 
Eq. (\ref{KK_MC}) combined with the
numerical estimate\cite{Kaasbjerg2016}
$M_{KK'}({\bf 0},{\bf 0}) = 145\, \text{eV} \mbox{\AA}^2/A$
gives $\Delta_{KK'}^{0000}=1.5$ meV for the 
intervalley impurity matrix element. 
This value is larger than the orbital
level spacing, in disagreement with the 
assumption Eq. (\ref{hierarchy}). 
Therefore, our results 
\eqref{eq:gxxmo} and \eqref{eq:rabimo}
 for the $g$-factor and Rabi frequency
can be applied for a Mo vacancy in MoS$_2$ only if
the distance of the vacancy from the QD centre 
is larger than 32 nm;
in this case the orbital level spacing 
exceeds the impurity matrix element, and hence
a perturbative treatment of the latter is valid. 
Of course, our results \eqref{eq:gxxmo} and \eqref{eq:rabimo}
can also be applied for other Mo-type defects
in MoS$_2$ or in other ML-TMDC materials, 
if the corresponding impurity matrix elements are below
the orbital level spacing. 

If the S-type impurity is a S vacancy, 
then the maximal value of intervalley impurity 
matrix element according to Eq.~(\ref{KK_XC}) is 
0.8 $\mu$eV, complying with the perturbative
hierarcy of Eq.~(\ref{hierarchy}). 
Then, the value of the maximal in-plane 
$g$ factor Eq.~\eqref{gmaxX} is 
$g_{xx,{\rm max}}^\textrm{S}=10^{-3}$,
and the maximal Rabi-frequency 
from Eq. \eqref{rabimaxX} is 
$\Omega_{\textrm{R,max}}^\textrm{S}=37$ MHz.
Note that in this example,
also the intravalley QD impurity matrix element is smaller
then the orbital level spacing, and therefore
using the harmonic-oscillator envelope functions as the orbital
basis is a reasonable approximation.
(Based on the numerical results of Ref.~\onlinecite{Kaasbjerg2016},
we estimate 
$M_{KK}({\bf 0},{\bf 0}) = 15\, \text{eV} \mbox{\AA}^2/A$,
implying
 $\Delta_{KK}^{0000}=150$ $\mu$eV  for the
parameter set considered here.)

\section{Conclusions}
\label{discussion}

In this work, we have proposed and analysed a way
to coherently control a single-electron spin-valley qubit, 
that is defined in a QD in a ML-TMDC material,
e.g., MoS$_2$.
Qubit control is performed in a fashion similar to electrically
driven spin resonance: 
a resonant ac electric field drives Rabi oscillations with
the help of an in-plane magnetic field and a
short-range impurity;
the former flips the spin and the latter flips the valley. 
For the case of a S-type impurity in the QD, 
we estimated that electrically driven qubit Rabi frequencies 
of the order of 10-100 MHz can be achieved. 
We also revealed and discussed the unconventional 
dependence of the
in-plane $g$ factor and the electrically driven Rabi frequency
on the impurity position, 
which arises for S-type impurities due
to symmetry-forbidden intervalley scattering. 
Our results provide design guidelines for efficient electric control
of spin-valley qubits in monolayer MoS$_2$,
which are promising building blocks for quantum information
processing experiments, expected to show boosted coherence
times, especially in isotopically purified samples.

\section{Acknowledgments}

We acknowledge useful discussions with 
G. Burkard, R. Frisenda, K. Kaasbjerg, J. Pet\H{o}, and L. Tapaszt\'o.
GSz and AP acknowledge the financial support of the
National Research, Development and Innovation Office of Hungary 
via the National Quantum Technologies Program NKP-2017-00001 
and the OTKA Grants 105149, 108676 and 124723. 
LC acknowledges funding from the European Union's 
Seventh Framework Programme (FP7/2007-2013) through 
the ERC Advanced Grant NOVGRAPHENE (GA No. 290846) 
and the Comunidad de Madrid through the grant MAD2D-CM, 
S2013/MIT-3007. AP is supported by the \'UNKP-17-4-III 
New National Excellence Program of the Ministry of 
Human Capacities of Hungary.

\appendix

\section{Notation and preliminaries: $k\cdot p$ perturbation theory}
\label{sec:notation}

Here, we review the elements of $k\cdot p$ theory
that are relevant for the
description of conduction-band electrons
in the ML-TMDC materials studied here.
These considerations are required to 
describe intervalley scattering
by S-type impurities, e.g., an S vacancy in MoS$_2$.

The single-electron Hamiltonian of the perfect crystal is
\bean\label{H0crystal}
\mathcal{H}_\text{cr}= \frac{p_x^2+p_y^2+p_z^2}{2m_{\rm e}} + \mathcal{V}_\text{cr}(\vec r,z),
\eean
where the crystal potential is denoted by  $\mathcal{V}_\text{cr}(\vec r,z)$ with 
$\vec r=(x,y)$.
The  eigenstates 
of $\mathcal H_\text{cr}$
are Bloch states:
\bean \label{Bloch}
\Psi_{\alpha, \vec k} (\vec r,z)= \frac{1}{\sqrt{A}} 
e^{i\vec k \vec r} u_{\alpha,\vec k} (\vec r,z),
\eean
where $\alpha$ is the band index,
$A$ is the sample area, 
and the lattice-periodic $u$ functions are normalized as
\bean
\int_{-\infty}^{\infty} dz 
\int_\text{sample} d^2 \vec r |u_{\alpha, \vec k}(\vec r,z)|^2 = A.
\eean
The corresponding eigenvalues (i.e., the band structure) are denoted 
as $\epsilon_{\alpha,\vec k}$.

Our quantum dot holds a single conduction electron 
in a spatially slowly varying 
confinement potential.
In this setup, the electronic wave function can be described
via Bloch states from the vicinity of the conduction-band
edge, that is,
from the vicinity of the $K$ and $K'$ points of the 
Brillouin zone. 
For crystal momenta around, e.g., $K$, 
we can 
use first-order perturbation theory to express
the conduction-band ($\alpha = c$) Bloch states 
$\Psi_{c,\vec K +\vec k}$
with the band-edge Bloch states
$\Psi_{\alpha,\vec K}$ -- 
this procedure is called $k\cdot p$ perturbation theory. 
In this theory, the so-called Kohn-Luttinger basis functions 
are used to form a matrix representation of the 
Hamiltonian:
\bean
\label{eq:kohnluttinger}
\phi_{\alpha, \vec K + \vec k} (\vec r,z) &=& 
e^{i\vec k \vec r} \Psi_{\alpha, \vec K}(\vec r,z)
\nonumber \\
&=&
\frac{1}{\sqrt{A}} e^{i\vec K \vec r}
e^{i\vec k \vec r}
u_{\alpha \vec K}(\vec r,z).
\eean
In the Kohn-Luttinger  basis, the crystal Hamiltonian
reads
\bean \label{KL_Hamilton}
\left(\mathcal{H}_\text{cr}\right)_{\alpha, \vec K+\vec k, \alpha', \vec K+ \vec k'} 
= \epsilon_{\alpha} \delta_{\alpha\alpha'} \delta_{\vec k \vec k'}
+ 
\delta_{\vec k \vec k'} \vec k \cdot
\vec P_{\alpha\alpha'},
\eean
where 
$\epsilon_\alpha = \epsilon_{\alpha,\vec K}$, 
the fully diagonal 
kinetic energy term $\hbar^2 k^2 / (2m_{\rm e})$ was omitted,
and
\bean\label{Pkdotp}
\vec P_{\alpha\alpha'} = \frac 1 A \frac{\hbar}{m_{\rm e}}
\braket{u_{\alpha\vec K}| \vec p | u_{\alpha'\vec K}}(1-\delta_{\alpha,\alpha'}).
\eean
with $\vec p = (p_x,p_y)$.

The next step in $k\cdot p$ theory is to reduce the multi-band problem
defined by the crystal Hamiltonian $\mathcal{H}_\text{cr}$
to the conduction band.
For this, second-order Schrieffer-Wolff
perturbation theory\cite{Winkler} can be applied,
where the fully diagonal first term
of Eq.~\eqref{KL_Hamilton} is the unperturbed Hamiltonian, 
and the second ($k\cdot p$) term of Eq.~\eqref{KL_Hamilton}
forms the perturbation.
The Schrieffer-Wolff transformation has the form 
$\tilde {\mathcal{H}}_\text{cr} = e^{-S} \mathcal{H}_\text{cr} e^{S}$,
where the expression for the 
anti-Hermitian
first-order transformation 
matrix is\cite{Winkler}
\bean
\label{eq:schriefferwolff}
S_{c\vec K + \vec k, \alpha \vec K + \vec k'} 
\approx 
- \delta_{\vec k \vec k'}
\frac{\vec k \cdot \vec P_{c\alpha}}{\epsilon_{c}-\epsilon_{\alpha}},
\eean
and the resulting effective conduction-band Hamiltonian 
reads
\bean
\left(\tilde{\mathcal{H}}_\text{cr}\right)_{c\vec K + \vec k,c\vec K + \vec k'} =
\vec k  \frac{\hbar^2}{2m^*} \vec k \,
\delta_{\vec k,\vec k'} 
\label{eq:hkp}
\eean
Here, 
we introduced the inverse effective mass tensor via
\bean
\frac 1 {m^*} = 
\frac{2}{\hbar^2}
\sum_{\alpha \neq c} 
\frac{ \vec P_{c \alpha} \circ \vec P_{\alpha c} }{\epsilon_{c} - \epsilon_{\alpha}}.
\eean
where $\circ$ is the dyadic product.
Note that since a threefold rotation around an out-of-plane
axis is a symmetry of the lattice, 
the effective mass is isotropic, i.e., this tensor is diagonal.

The effective conduction-band Hamiltonian matrix 
$\tilde{\mathcal H}_\text{cr}$ is diagonal, so 
its eigenvectors $c^{(\vec k)}$ can
be labelled by the wave vector $\vec k$,
and have the trivial structure
$c^{(\vec k)}_{\vec k'} = \delta_{\vec k,\vec k'}$.
The corresponding eigenvalue is $\hbar k^2/(2m^*)$. 
To obtain the corresponding Bloch state
$\Psi_{c,\vec K + \vec k}(\vec r,z)$, we have to invert the 
Schrieffer-Wolff transformation, 
\bean
\label{eq:swinverse}
\Psi_{c,\vec K + \vec k}&=& 
\sum_{\alpha \vec k'}
	\left[
		\sum_{\vec k'' }
		\left(	e^S \right)_{\alpha,\vec K + \vec k',c \vec K + \vec k''} 
			c^{(\vec k)}_{\vec k''}
	\right]
\phi_{\alpha,\vec K + \vec k'}. \nonumber
\\
\eean
Making a linear expansion of the exponential, and
using the 
the first-order result \eqref{eq:schriefferwolff} 
for the transformation matrix, we obtain
the perturbative first-order approximation
$\Psi^{(1)}_{c,\vec K + \vec k}$
for the Bloch state
$\Psi_{c,\vec K + \vec k}$
as
\bean
\Psi^{(1)}_{c,\vec K + \vec k} &=& 
\phi_{c,\vec K + \vec k}+
\sum_{\alpha \neq c}
\frac{\vec k \cdot \vec P_{\alpha c}}{\epsilon_{c } - \epsilon_{\alpha}} \phi_{\alpha, \vec K + \vec k}
\label{eq:perturbativebloch}
\\
&=&
e^{i\vec k \vec r}
\left(
\Psi_{c,\vec K}+
\sum_{\alpha \neq c}
\frac{\vec k \cdot \vec P_{\alpha c}}{\epsilon_{c} - \epsilon_{\alpha }} \Psi_{\alpha, \vec K}
\right), \label{eq:kpresult}
\eean
where the arguments $\vec r,z$ are suppressed.
This result is used in this work to evaluate the 
linear-in-momentum component of the intervalley matrix 
elements of a short-range impurity. 
Note that this treatment, i.e., taking into account
the $k\cdot p$-induced hybridisation of the conduction band
with remote bands, is required only for S-type impurities,
where 
a linear-in-$\vec k$ expansion of the intervalley 
matrix element is needed to obtain a nonzero result.

\section{S-type impurity: symmetry constraints of the
bulk intervalley matrix element}
\label{app:symmetry}

For an S-type impurity, the
symmetries of the 
conduction-band-edge Bloch function 
and the atomic structure 
pose the constraint
\eqref{eq:sapprox}
on the linear-in-$\vec k$ contribution of the bulk intervalley
impurity matrix element $M_{KK'}(\vec k,\vec k')$,
 defined in Eq.~\eqref{Mkk}.
 Here we derive this constraint, with the specification
 that the Bloch states in the definition \eqref{Mkk}
 are given by the $k\cdot p$ result \eqref{eq:kpresult}. 

The generic expression for the linear-in-$\vec k$
bulk intervalley impurity matrix element,
without any symmetry restrictions, 
is
\bean
\label{eq:mkkgeneral}
M_{KK'}(\vec k,\vec k') = 
z_0 + z_x k_x + z_y k_y + 
z'_x k'_x + z'_y k'_y,
\eean
where the 5 coefficients are complex numbers. 
It is natural to expect that the rotational symmetry ($C_3$) 
provides a relation between 
$M_{KK'}(C_3 \vec k,C_3\vec k')$ and
$M_{KK'}(\vec k,\vec k')$,
and that time-reversal symmetry provides 
a relation between 
$M_{KK'}(-\vec k',-\vec k)$ and
$M_{KK'}(\vec k,\vec k')$.
We derive these relations here, transform them 
to relations between the 5 $z$ coefficients,
and thereby simplify the general expression in
Eq.~\eqref{eq:mkkgeneral}.

To exploit the rotational symmetry $C_3$, 
we start from
\bean
\label{eq:mkkc3}
M_{KK'}(C_3 \vec k, C_3\vec k') = 
\braket{
	\Psi^{(1)}_{c,\vec K+C_3 \vec k} | 
	\mathcal{U}_\text{imp} | 
	\Psi^{(1)}_{c,\vec K'+C_3  \vec k'}
}.
\eean
On the one hand, from Eq.~\eqref{eq:kpresult}, we know
\begin{eqnarray}
\label{eq:rotation1}
\Psi^{(1)}_{c,\vec K + C_3 \vec k} 
= e^{i (C_3 \vec k) \vec r}
\left[
\Psi_{c,\vec K}+
\sum_{\alpha \neq c}
\frac{(C_3 \vec k) \cdot \vec P_{\alpha c}}{\epsilon_{c \vec K} - \epsilon_{\alpha \vec K}} \Psi_{\alpha, \vec K}
\right],\nonumber\\
\end{eqnarray}
and that this is a non-degenerate 
Bloch state with energy eigenvalue $\epsilon_{c,\vec K + \vec k}$.
On the other hand, we know\cite{GuiBinLiu_intervalley,Kaasbjerg2016} 
how the band-edge Bloch 
states transform under rotation around an 
S site: 
$C_3 \Psi_{\alpha,\vec K} = \omega^{n_\alpha} \Psi_{\alpha, \vec K}$
with 
$\omega = e^{i2\pi/3}$ and
$n_\alpha \in \{-1,0,1\}$, 
and $n_c = 1$ in particular.
Combining these with Eq.~\eqref{eq:kpresult}, we find
\begin{eqnarray}
\label{eq:rotation2}
C_3 \Psi^{(1)}_{c,\vec K+ \vec k} = 
e^{i (C_3 \vec k) \vec r}
\left[
\omega \Psi_{c,\vec K}+
\sum_{\alpha \neq c}
\frac{\vec k \cdot \vec P_{\alpha c}}{\epsilon_{c \vec K} - \epsilon_{\alpha \vec K}} \omega^{n_\alpha}\Psi_{\alpha, \vec K}
\right].\nonumber\\
\end{eqnarray}
Importantly, this is also a Bloch state with energy 
$\epsilon_{c,\vec K + \vec k}$.
Since the Bloch states 
\eqref{eq:rotation1} and \eqref{eq:rotation2}
have the same energy 
and momentum, they must 
be identical, up to a complex phase factor.
This requirement implies
\bean
\label{eq:omegaK}
\Psi^{(1)}_{c,\vec K + C_3 \vec k} =
\omega^{-1} C_3 \Psi^{(1)}_{c,\vec K + \vec k}.
\eean

We  describe the $K'$-valley Bloch states
as time-reversed $K$-valley Bloch states,
$\Psi_{\alpha,\vec K'} = \Psi_{\alpha,\vec K}^*$, 
yielding 
$C_3 \Psi_{\alpha,\vec K'} = \omega^{-n_\alpha} \Psi_{\alpha,\vec K'}$,
and 
\bean
\label{eq:omegaKprime}
\Psi^{(1)}_{c,\vec K' + C_3 \vec k'} =
\omega C_3 \Psi^{(1)}_{c,\vec K' + \vec k'}.
\eean
Combining Eqs.~\eqref{eq:omegaK}
and \eqref{eq:omegaKprime} in 
Eq.~\eqref{eq:mkkc3}, 
and exploiting the rotational symmetry 
$C_3 \mathcal{U}_\text{imp} C_3^{-1} = \mathcal{U}_\text{imp}$
of the 
impurity potential, 
results in 
\bean
\label{eq:mkkrot}
M_{KK'} (C_3 \vec k, C_3\vec k') = 
\omega^{-1} M_{KK'} (\vec k, \vec k').
\eean

To exploit time reversal symmetry $T$, which is the complex
conjugation in our case, fulfilling $T = T^{-1}$, 
we start from
\bean
\label{eq:mkktrs}
M_{KK'}(-\vec k',-\vec k) =
\braket{
	\Psi^{(1)}_{c,\vec K - \vec k'} |
	\mathcal{U}_\text{imp} |
	\Psi^{(1)}_{c,\vec K' - \vec k}
}.
\eean
Again, we describe the $K'$-valley Bloch states as
time-reversed $K$-valley Bloch states:
$\Psi^{(1)}_{c,\vec K'-\vec k} = 
\left[\Psi^{(1)}_{c,\vec K + \vec k}\right]^*$.
Using this together with the time-reversal symmetry
$T\mathcal{U}_\text{imp}T^{-1} = \mathcal{U}_\text{imp}$ of the impurity potential, 
i.e., that the latter is real-valued,
we find
\bean
M_{KK'}(-\vec k',-\vec k) = M_{KK'} (\vec k, \vec k'). 
\eean

The symmetry-based  equations 
\eqref{eq:mkkrot} and \eqref{eq:mkktrs}
must hold for any values of the 
wave-vector components $k_x$, $k_y$, $k'_x$, $k'_y$. 
Therefore, these equations form a set of 
10 complex linear equations for the 5 complex unknown
$z$ coefficients.
We solve this linear set, and apply the 
resulting relations in Eq.~\eqref{eq:mkkgeneral}, yielding
\bean
M_{KK'} (\vec k, \vec k') =
(k_x - i k_y - k'_x + i k'_y) z_x.
\eean
Clearly, $z_x$ inherits a global complex phase factor from the 
band-edge Bloch state $\Psi_{c,\vec K}$, and hence
with appropriate choice of global complex phase 
for  $\Psi_{c,\vec K}$ we can make
$z_x$ positive.
Hence, we have proven that the form of the bulk intervalley 
impurity matrix elements follow Eq.~\eqref{eq:sapprox},
with the identification $z_x = \gamma/A$.

Note that the key difference between the Mo-type and
S-type impurities is the transformation factor $n_c$. 
I.e., both types of impurities preserve the $C_3$
rotation around the impurity site as a symmetry of the structure.
However, the rotation around the Mo-type impurity
transforms\cite{Kaasbjerg2016} the conduction-band-edge state
$\Psi^{(1)}_{c,\vec K}$
as
$C_3 \Psi^{(1)}_{c,\vec K} = \Psi^{(1)}_{c,\vec K}$, that is, 
with $n_c = 0$, implying that 
$M_{KK'}({\bf 0},{\bf 0})$ is finite in general for a Mo-type
impurity, and therefore can be used 
to approximate the intervalley impurity
matrix elements between the QD wave functions.

\section{Quantum-dot wave functions from the envelope-function
approximation}
\label{app:envelopefunction}

We have described the wave function of the
electron in our QD via Eq.~\eqref{nmvs_function},
i.e., as a wave packet, formed by the conduction-band Bloch 
states $\Psi^{(1)}_{c, \vec k}$ expressed from 
$k\cdot p$ theory.
Here we summarize how Eq.~\eqref{nmvs_function}
is obtained from the envelope-function approximation (EFA).

The Hamiltonian
$\mathcal{H} = \mathcal{H}_\text{cr}+
\mathcal{V}_\text{conf}$
of the electron in the QD is the
sum of the crystal Hamiltonian $\mathcal{H}_\text{cr}$
and the QD confinement potential $\mathcal{V}_\text{conf}$.
The 
EFA is an extension of $k\cdot p$ theory: it also relies
on the Kohn-Luttinger matrix representation of 
the Hamiltonian, which is now given by 
Eq.~\eqref{KL_Hamilton}
plus 
$\left(\mathcal{V}_\text{conf}\right)_{\alpha\vec k,\alpha' \vec k'}$.
Exploiting the spatially slowly varying nature of 
$\mathcal{V}_\text{conf}$ and
the orthogonality of the $u_{\alpha \vec K}$ 
functions of different bands,
the interband matrix elements are 
neglected, 
$\left(\mathcal{V}_\text{conf}\right)_{\alpha\vec k, \alpha' \vec k'} = 
 \delta_{\alpha\alpha'}\left( \mathcal{V}_\text{conf}\right)^{(\alpha)}_{\vec k \vec k'}$.
Furthermore, intervalley matrix elements of $\mathcal{V}_\text{conf}$ are neglected.
This approximation is justified by the
numerical results of Ref.~\onlinecite{GuiBinLiu_intervalley},
which imply that in our example, the order of magnitude of 
the confinement-induced intervalley matrix element does exceed 1 neV,
it remains well below the impurity-induced intervalley matrix 
element of a single S vacancy ($\sim 1 \,\mu\text{eV}$, see our
Section \ref{sec:mos2}).

%

Similarly to $k\cdot p$ theory, the next step in the EFA 
is to reduce the multi-band problem 
of $\mathcal{H}$ 
to the conduction band via the Schrieffer-Wolff transformation,
treating the confinement potential as part of the perturbation. 
Importantly, since the interband matrix elements of the 
confinement potential are neglected, the interband
perturbation remains the  same as in the 
absence of the confinement potential, 
which implies that the Schrieffer-Wolff transformation 
matrix $S$ has exactly the same form as in  $k\cdot p$ theory, 
i.e., Eq.~\eqref{eq:schriefferwolff}.
The transformation results in 
the following effective conduction-band Hamiltonian 
(cf.~Eq.~\eqref{eq:hkp}):
\bean
\tilde{\mathcal{H}}_{c\vec K + \vec k,c\vec K + \vec k'} =
  \frac{\hbar^2 k^2}{2m^*} \,
\delta_{\vec k,\vec k'} 
+
\left(
\mathcal{V}_\text{conf}
\right)^{(c)}_{\vec K + \vec k,\vec K+ \vec k'}.
\label{eq:hefkspace}
\eean
Recall that the effective mass tensor is isotropic; 
this fact was used in Eqs.~\eqref{eq:vconf} and \eqref{envelope}.

Solutions $(c,E)$ of the matrix Schr\"odinger equation
\bean
\sum_{\vec k'} \tilde{\mathcal{H}}_{c\vec K + \vec k,c\vec K + \vec k'} 
c_{\vec k'}
= E c_{\vec k}
\eean
are usually derived by 
Fourier transforming this equation to a real-space
envelope-function Schr\"odinger equation,
yielding Eq.~\eqref{envelope},
finding the solution $(\Phi,E)$ of 
the latter, and Fourier transforming the envelope function $\Phi(x,y)$
to momentum space, yielding the coefficients $c_{\vec k}$.

To express the real-space wave function $\psi(\vec r,z)$
corresponding to the eigenvector $c$ of the 
transformed Hamiltonian $\tilde{\mathcal{H}}$, 
we have to invert the Schrieffer-Wolff transformation
[cf.~Eq.~\eqref{eq:swinverse}]:
\bean
\psi &=& 
\sum_{\alpha,\vec k}
	\left[
		\sum_{\vec k' }
		\left(	e^S \right)_{\alpha,\vec K + \vec k,c \vec K + \vec k'} 
			c_{\vec k'}
	\right]
\phi_{\alpha,\vec K + \vec k}.
\eean
Using a linear expansion of the
Schrieffer-Wolff transformation matrix in the perturbation, 
we obtain
\bean
\psi = \sum_{\vec k} c_{\vec k}
\Psi^{(1)}_{c,\vec K + \vec k},
\label{eq:realspacewfn}
\eean
which is then utilized in the main text as Eq.~\eqref{nmvs_function}.
In words, Eq.~\eqref{eq:realspacewfn}
expresses the fact that each 
electronic real-space energy eigenfunction
of the QD is a  packet of Bloch waves from the 
vicinity of the $K$ point,
and the coefficients $c_{\vec k}$ of this wave packet
can be obtained from the envelope-function
Schr\"odinger equation.

\section{The complete third-order qubit Hamiltonian}

In the main text, we have presented the effective spin-valley
qubit Hamiltonian that is required to describe the in-plane
$g$ factor and the electrically driven Rabi oscillations. 
For completeness, here we present the  
third-order effective Hamiltonian, including the
third-order time-independent term that was not
discussed in the main text.

Recall that performing a second-order (time-independent)
Schrieffer-Wolff transformation yields 
$H_\text{q} = H^{(1)}_\text{q} + H^{(2)}_\text{q}$ as
shown by Eqs.~\eqref{eq:h2bparallel}, 
\eqref{eq:hq1}, and
\eqref{eq:hq2}.
Performing a third-order time-dependent 
Schrieffer-Wolff transformation\cite{Romhanyi}
we find
\begin{equation}
\label{eq:tdswresult}
H_\textrm{q} = 
H_\textrm{q}^{(1)}
+ H^{(2)}_\textrm{q}
+ H^{(3)}_\textrm{q}
+ H^{(3)}_\textrm{q}(t),
\end{equation}
where
the first- and second-order terms are the
same as above, 
the time-dependent third-order term is given 
in Eq.~\eqref{eq:hqubit3}, 
and the time-independent third-order term is 
\begin{widetext}
\begin{eqnarray}\label{eq:h3e}
H_\textrm{q}^{(3)} &=&
 \frac 1 4 \mu_B^3 g_s^3 
\frac{B_\perp^2 B_z}{\Delta_{\textrm{SO}}^2} \sigma_z 
-
g_v \mu_B B_z\sum_{nm}\frac{\left(\Delta_{KK'}^{00nm}\right)^2}{(\Delta_{\textrm{SO}}+(n+m)\hbar \omega_0)^2}\sigma_z\nonumber\\
&+&2 \frac{g_s \mu_B B_\perp }{\Delta_{\textrm{SO}} \hbar \omega_0} \sum_{\substack{nm \\ (nm)\neq(00)}}\frac{1}{n+m}
\left[
\Delta^{00nm}_{KK'} \Delta^{00nm}_{KK}
\right] \sigma(\varphi_{KK'}^{00nm}+\varphi_B).
\end{eqnarray}
\end{widetext}
As in the main text, 
$\sigma_x$, $\sigma_y$, $\sigma_z$ are Pauli matrices
acting on the perturbed states 
$\ket{\Uparrow}$ and $\ket{\Downarrow}$, 
see Sec.~\ref{Spin-valleyQubit}. 
The three subsequent terms describe the 
nonlinear Zeeman effect, 
and a higher-order correction of the out-of-plane
and in-plane $g$-factors, respectively.

\bibliography{mos2quantumdot}

\begin{thebibliography}{51}%
\makeatletter
\providecommand \@ifxundefined [1]{%
 \@ifx{#1\undefined}
}%
\providecommand \@ifnum [1]{%
 \ifnum #1\expandafter \@firstoftwo
 \else \expandafter \@secondoftwo
 \fi
}%
\providecommand \@ifx [1]{%
 \ifx #1\expandafter \@firstoftwo
 \else \expandafter \@secondoftwo
 \fi
}%
\providecommand \natexlab [1]{#1}%
\providecommand \enquote  [1]{``#1''}%
\providecommand \bibnamefont  [1]{#1}%
\providecommand \bibfnamefont [1]{#1}%
\providecommand \citenamefont [1]{#1}%
\providecommand \href@noop [0]{\@secondoftwo}%
\providecommand \href [0]{\begingroup \@sanitize@url \@href}%
\providecommand \@href[1]{\@@startlink{#1}\@@href}%
\providecommand \@@href[1]{\endgroup#1\@@endlink}%
\providecommand \@sanitize@url [0]{\catcode `\\12\catcode `\$12\catcode
  `\&12\catcode `\#12\catcode `\^12\catcode `\_12\catcode `\%12\relax}%
\providecommand \@@startlink[1]{}%
\providecommand \@@endlink[0]{}%
\providecommand \url  [0]{\begingroup\@sanitize@url \@url }%
\providecommand \@url [1]{\endgroup\@href {#1}{\urlprefix }}%
\providecommand \urlprefix  [0]{URL }%
\providecommand \Eprint [0]{\href }%
\providecommand \doibase [0]{http://dx.doi.org/}%
\providecommand \selectlanguage [0]{\@gobble}%
\providecommand \bibinfo  [0]{\@secondoftwo}%
\providecommand \bibfield  [0]{\@secondoftwo}%
\providecommand \translation [1]{[#1]}%
\providecommand \BibitemOpen [0]{}%
\providecommand \bibitemStop [0]{}%
\providecommand \bibitemNoStop [0]{.\EOS\space}%
\providecommand \EOS [0]{\spacefactor3000\relax}%
\providecommand \BibitemShut  [1]{\csname bibitem#1\endcsname}%
\let\auto@bib@innerbib\@empty
\bibitem [{\citenamefont {Hanson}\ \emph {et~al.}(2007)\citenamefont {Hanson},
  \citenamefont {Kouwenhoven}, \citenamefont {Petta}, \citenamefont {Tarucha},\
  and\ \citenamefont {Vandersypen}}]{Hanson-rmp}%
  \BibitemOpen
  \bibfield  {author} {\bibinfo {author} {\bibfnamefont {R.}~\bibnamefont
  {Hanson}}, \bibinfo {author} {\bibfnamefont {L.~P.}\ \bibnamefont
  {Kouwenhoven}}, \bibinfo {author} {\bibfnamefont {J.~R.}\ \bibnamefont
  {Petta}}, \bibinfo {author} {\bibfnamefont {S.}~\bibnamefont {Tarucha}}, \
  and\ \bibinfo {author} {\bibfnamefont {L.~M.~K.}\ \bibnamefont
  {Vandersypen}},\ }\href@noop {} {\bibfield  {journal} {\bibinfo  {journal}
  {Rev. Mod. Phys.}\ }\textbf {\bibinfo {volume} {79}},\ \bibinfo {pages}
  {1217} (\bibinfo {year} {2007})}\BibitemShut {NoStop}%
\bibitem [{\citenamefont {Petta}\ \emph {et~al.}(2005)\citenamefont {Petta},
  \citenamefont {Johnson}, \citenamefont {Taylor}, \citenamefont {Laird},
  \citenamefont {Yacoby}, \citenamefont {Lukin}, \citenamefont {Marcus},
  \citenamefont {Hanson},\ and\ \citenamefont {Gossard}}]{Petta}%
  \BibitemOpen
  \bibfield  {author} {\bibinfo {author} {\bibfnamefont {J.~R.}\ \bibnamefont
  {Petta}}, \bibinfo {author} {\bibfnamefont {A.~C.}\ \bibnamefont {Johnson}},
  \bibinfo {author} {\bibfnamefont {J.~M.}\ \bibnamefont {Taylor}}, \bibinfo
  {author} {\bibfnamefont {E.~A.}\ \bibnamefont {Laird}}, \bibinfo {author}
  {\bibfnamefont {A.}~\bibnamefont {Yacoby}}, \bibinfo {author} {\bibfnamefont
  {M.~D.}\ \bibnamefont {Lukin}}, \bibinfo {author} {\bibfnamefont {C.~M.}\
  \bibnamefont {Marcus}}, \bibinfo {author} {\bibfnamefont {M.~P.}\
  \bibnamefont {Hanson}}, \ and\ \bibinfo {author} {\bibfnamefont {A.~C.}\
  \bibnamefont {Gossard}},\ }\href {\doibase 10.1126/science.1116955}
  {\bibfield  {journal} {\bibinfo  {journal} {Science}\ }\textbf {\bibinfo
  {volume} {309}},\ \bibinfo {pages} {2180} (\bibinfo {year}
  {2005})}\BibitemShut {NoStop}%
\bibitem [{\citenamefont {Flindt}\ \emph {et~al.}(2006)\citenamefont {Flindt},
  \citenamefont {S\o{}rensen},\ and\ \citenamefont {Flensberg}}]{Flindt}%
  \BibitemOpen
  \bibfield  {author} {\bibinfo {author} {\bibfnamefont {C.}~\bibnamefont
  {Flindt}}, \bibinfo {author} {\bibfnamefont {A.~S.}\ \bibnamefont
  {S\o{}rensen}}, \ and\ \bibinfo {author} {\bibfnamefont {K.}~\bibnamefont
  {Flensberg}},\ }\href {\doibase 10.1103/PhysRevLett.97.240501} {\bibfield
  {journal} {\bibinfo  {journal} {Phys. Rev. Lett.}\ }\textbf {\bibinfo
  {volume} {97}},\ \bibinfo {pages} {240501} (\bibinfo {year}
  {2006})}\BibitemShut {NoStop}%
\bibitem [{\citenamefont {Golovach}\ \emph {et~al.}(2006)\citenamefont
  {Golovach}, \citenamefont {Borhani},\ and\ \citenamefont {Loss}}]{Golovach}%
  \BibitemOpen
  \bibfield  {author} {\bibinfo {author} {\bibfnamefont {V.~N.}\ \bibnamefont
  {Golovach}}, \bibinfo {author} {\bibfnamefont {M.}~\bibnamefont {Borhani}}, \
  and\ \bibinfo {author} {\bibfnamefont {D.}~\bibnamefont {Loss}},\ }\href
  {\doibase 10.1103/PhysRevB.74.165319} {\bibfield  {journal} {\bibinfo
  {journal} {Phys. Rev. B}\ }\textbf {\bibinfo {volume} {74}},\ \bibinfo
  {pages} {165319} (\bibinfo {year} {2006})}\BibitemShut {NoStop}%
\bibitem [{\citenamefont {Rashba}(2008)}]{Rashba2008}%
  \BibitemOpen
  \bibfield  {author} {\bibinfo {author} {\bibfnamefont {E.~I.}\ \bibnamefont
  {Rashba}},\ }\href {\doibase 10.1103/PhysRevB.78.195302} {\bibfield
  {journal} {\bibinfo  {journal} {Phys. Rev. B}\ }\textbf {\bibinfo {volume}
  {78}},\ \bibinfo {pages} {195302} (\bibinfo {year} {2008})}\BibitemShut
  {NoStop}%
\bibitem [{\citenamefont {Nowack}\ \emph {et~al.}(2007)\citenamefont {Nowack},
  \citenamefont {Koppens}, \citenamefont {Nazarov},\ and\ \citenamefont
  {Vandersypen}}]{Nowack-esr}%
  \BibitemOpen
  \bibfield  {author} {\bibinfo {author} {\bibfnamefont {K.~C.}\ \bibnamefont
  {Nowack}}, \bibinfo {author} {\bibfnamefont {F.~H.~L.}\ \bibnamefont
  {Koppens}}, \bibinfo {author} {\bibfnamefont {Y.~V.}\ \bibnamefont
  {Nazarov}}, \ and\ \bibinfo {author} {\bibfnamefont {L.~M.~K.}\ \bibnamefont
  {Vandersypen}},\ }\href {\doibase 10.1126/science.1148092} {\bibfield
  {journal} {\bibinfo  {journal} {Science}\ }\textbf {\bibinfo {volume}
  {318}},\ \bibinfo {pages} {1430} (\bibinfo {year} {2007})}\BibitemShut
  {NoStop}%
\bibitem [{\citenamefont {Nadj-Perge}\ \emph {et~al.}(2010)\citenamefont
  {Nadj-Perge}, \citenamefont {Frolov}, \citenamefont {Bakkers},\ and\
  \citenamefont {Kouwenhoven}}]{Nadj-Perge2010}%
  \BibitemOpen
  \bibfield  {author} {\bibinfo {author} {\bibfnamefont {S.}~\bibnamefont
  {Nadj-Perge}}, \bibinfo {author} {\bibfnamefont {S.~M.}\ \bibnamefont
  {Frolov}}, \bibinfo {author} {\bibfnamefont {E.~P. A.~M.}\ \bibnamefont
  {Bakkers}}, \ and\ \bibinfo {author} {\bibfnamefont {L.~P.}\ \bibnamefont
  {Kouwenhoven}},\ }\href@noop {} {\bibfield  {journal} {\bibinfo  {journal}
  {Nature}\ }\textbf {\bibinfo {volume} {468}},\ \bibinfo {pages} {1084}
  (\bibinfo {year} {2010})}\BibitemShut {NoStop}%
\bibitem [{\citenamefont {Nadj-Perge}\ \emph {et~al.}(2012)\citenamefont
  {Nadj-Perge}, \citenamefont {Pribiag}, \citenamefont {van~den Berg},
  \citenamefont {Zuo}, \citenamefont {Plissard}, \citenamefont {Bakkers},
  \citenamefont {Frolov},\ and\ \citenamefont {Kouwenhoven}}]{NadjPerge}%
  \BibitemOpen
  \bibfield  {author} {\bibinfo {author} {\bibfnamefont {S.}~\bibnamefont
  {Nadj-Perge}}, \bibinfo {author} {\bibfnamefont {V.~S.}\ \bibnamefont
  {Pribiag}}, \bibinfo {author} {\bibfnamefont {J.~W.~G.}\ \bibnamefont
  {van~den Berg}}, \bibinfo {author} {\bibfnamefont {K.}~\bibnamefont {Zuo}},
  \bibinfo {author} {\bibfnamefont {S.~R.}\ \bibnamefont {Plissard}}, \bibinfo
  {author} {\bibfnamefont {E.~P. A.~M.}\ \bibnamefont {Bakkers}}, \bibinfo
  {author} {\bibfnamefont {S.~M.}\ \bibnamefont {Frolov}}, \ and\ \bibinfo
  {author} {\bibfnamefont {L.~P.}\ \bibnamefont {Kouwenhoven}},\ }\href
  {\doibase 10.1103/PhysRevLett.108.166801} {\bibfield  {journal} {\bibinfo
  {journal} {Phys. Rev. Lett.}\ }\textbf {\bibinfo {volume} {108}},\ \bibinfo
  {pages} {166801} (\bibinfo {year} {2012})}\BibitemShut {NoStop}%
\bibitem [{\citenamefont {Nebel}(2013)}]{Nebel}%
  \BibitemOpen
  \bibfield  {author} {\bibinfo {author} {\bibfnamefont {C.~E.}\ \bibnamefont
  {Nebel}},\ }\href@noop {} {\bibfield  {journal} {\bibinfo  {journal} {Nat.
  Mater.}\ }\textbf {\bibinfo {volume} {12}},\ \bibinfo {pages} {690} (\bibinfo
  {year} {2013})}\BibitemShut {NoStop}%
\bibitem [{\citenamefont {Maurand}\ \emph {et~al.}(2016)\citenamefont
  {Maurand}, \citenamefont {Jehl}, \citenamefont {Kotekar-Patil}, \citenamefont
  {Corna}, \citenamefont {Bohuslavskyi}, \citenamefont {Lavi{\'e}ville},
  \citenamefont {Hutin}, \citenamefont {Barraud}, \citenamefont {Vinet},
  \citenamefont {Sanquer},\ and\ \citenamefont {De~Franceschi}}]{Maurand}%
  \BibitemOpen
  \bibfield  {author} {\bibinfo {author} {\bibfnamefont {R.}~\bibnamefont
  {Maurand}}, \bibinfo {author} {\bibfnamefont {X.}~\bibnamefont {Jehl}},
  \bibinfo {author} {\bibfnamefont {D.}~\bibnamefont {Kotekar-Patil}}, \bibinfo
  {author} {\bibfnamefont {A.}~\bibnamefont {Corna}}, \bibinfo {author}
  {\bibfnamefont {H.}~\bibnamefont {Bohuslavskyi}}, \bibinfo {author}
  {\bibfnamefont {R.}~\bibnamefont {Lavi{\'e}ville}}, \bibinfo {author}
  {\bibfnamefont {L.}~\bibnamefont {Hutin}}, \bibinfo {author} {\bibfnamefont
  {S.}~\bibnamefont {Barraud}}, \bibinfo {author} {\bibfnamefont
  {M.}~\bibnamefont {Vinet}}, \bibinfo {author} {\bibfnamefont
  {M.}~\bibnamefont {Sanquer}}, \ and\ \bibinfo {author} {\bibfnamefont
  {S.}~\bibnamefont {De~Franceschi}},\ }\href@noop {} {\bibfield  {journal}
  {\bibinfo  {journal} {Nat. Comm.}\ }\textbf {\bibinfo {volume} {7}},\
  \bibinfo {pages} {13575} (\bibinfo {year} {2016})}\BibitemShut {NoStop}%
\bibitem [{\citenamefont {Zwanenburg}\ \emph {et~al.}(2013)\citenamefont
  {Zwanenburg}, \citenamefont {Dzurak}, \citenamefont {Morello}, \citenamefont
  {Simmons}, \citenamefont {Hollenberg}, \citenamefont {Klimeck}, \citenamefont
  {Rogge}, \citenamefont {Coppersmith},\ and\ \citenamefont
  {Eriksson}}]{Zwanenburg}%
  \BibitemOpen
  \bibfield  {author} {\bibinfo {author} {\bibfnamefont {F.~A.}\ \bibnamefont
  {Zwanenburg}}, \bibinfo {author} {\bibfnamefont {A.~S.}\ \bibnamefont
  {Dzurak}}, \bibinfo {author} {\bibfnamefont {A.}~\bibnamefont {Morello}},
  \bibinfo {author} {\bibfnamefont {M.~Y.}\ \bibnamefont {Simmons}}, \bibinfo
  {author} {\bibfnamefont {L.~C.~L.}\ \bibnamefont {Hollenberg}}, \bibinfo
  {author} {\bibfnamefont {G.}~\bibnamefont {Klimeck}}, \bibinfo {author}
  {\bibfnamefont {S.}~\bibnamefont {Rogge}}, \bibinfo {author} {\bibfnamefont
  {S.~N.}\ \bibnamefont {Coppersmith}}, \ and\ \bibinfo {author} {\bibfnamefont
  {M.~A.}\ \bibnamefont {Eriksson}},\ }\href@noop {} {\bibfield  {journal}
  {\bibinfo  {journal} {Rev. Mod. Phys.}\ }\textbf {\bibinfo {volume} {85}},\
  \bibinfo {pages} {961} (\bibinfo {year} {2013})}\BibitemShut {NoStop}%
\bibitem [{\citenamefont {Culcer}\ \emph {et~al.}(2012)\citenamefont {Culcer},
  \citenamefont {Saraiva}, \citenamefont {Koiller}, \citenamefont {Hu},\ and\
  \citenamefont {Das~Sarma}}]{Culcer}%
  \BibitemOpen
  \bibfield  {author} {\bibinfo {author} {\bibfnamefont {D.}~\bibnamefont
  {Culcer}}, \bibinfo {author} {\bibfnamefont {A.~L.}\ \bibnamefont {Saraiva}},
  \bibinfo {author} {\bibfnamefont {B.}~\bibnamefont {Koiller}}, \bibinfo
  {author} {\bibfnamefont {X.}~\bibnamefont {Hu}}, \ and\ \bibinfo {author}
  {\bibfnamefont {S.}~\bibnamefont {Das~Sarma}},\ }\href@noop {} {\bibfield
  {journal} {\bibinfo  {journal} {Phys. Rev. Lett.}\ }\textbf {\bibinfo
  {volume} {108}},\ \bibinfo {pages} {126804} (\bibinfo {year}
  {2012})}\BibitemShut {NoStop}%
\bibitem [{\citenamefont {Rohling}\ and\ \citenamefont
  {Burkard}(2012)}]{Rohling}%
  \BibitemOpen
  \bibfield  {author} {\bibinfo {author} {\bibfnamefont {N.}~\bibnamefont
  {Rohling}}\ and\ \bibinfo {author} {\bibfnamefont {G.}~\bibnamefont
  {Burkard}},\ }\href@noop {} {\bibfield  {journal} {\bibinfo  {journal} {New
  Journal of Physics}\ }\textbf {\bibinfo {volume} {14}},\ \bibinfo {pages}
  {083008} (\bibinfo {year} {2012})}\BibitemShut {NoStop}%
\bibitem [{\citenamefont {Laird}\ \emph {et~al.}(2013)\citenamefont {Laird},
  \citenamefont {Pei},\ and\ \citenamefont {Kouwenhoven}}]{Laird}%
  \BibitemOpen
  \bibfield  {author} {\bibinfo {author} {\bibfnamefont {E.~A.}\ \bibnamefont
  {Laird}}, \bibinfo {author} {\bibfnamefont {F.}~\bibnamefont {Pei}}, \ and\
  \bibinfo {author} {\bibfnamefont {L.~P.}\ \bibnamefont {Kouwenhoven}},\
  }\href@noop {} {\bibfield  {journal} {\bibinfo  {journal} {Nat. Nanotech.}\
  }\textbf {\bibinfo {volume} {8}},\ \bibinfo {pages} {565} (\bibinfo {year}
  {2013})}\BibitemShut {NoStop}%
\bibitem [{\citenamefont {P\'alyi}\ and\ \citenamefont
  {Burkard}(2011)}]{Palyi-valley-resonance}%
  \BibitemOpen
  \bibfield  {author} {\bibinfo {author} {\bibfnamefont {A.}~\bibnamefont
  {P\'alyi}}\ and\ \bibinfo {author} {\bibfnamefont {G.}~\bibnamefont
  {Burkard}},\ }\href {\doibase 10.1103/PhysRevLett.106.086801} {\bibfield
  {journal} {\bibinfo  {journal} {Phys. Rev. Lett.}\ }\textbf {\bibinfo
  {volume} {106}},\ \bibinfo {pages} {086801} (\bibinfo {year}
  {2011})}\BibitemShut {NoStop}%
\bibitem [{\citenamefont {P\'{a}lyi}\ and\ \citenamefont
  {Burkard}(2010)}]{Palyi-cnt-spinblockade}%
  \BibitemOpen
  \bibfield  {author} {\bibinfo {author} {\bibfnamefont {A.}~\bibnamefont
  {P\'{a}lyi}}\ and\ \bibinfo {author} {\bibfnamefont {G.}~\bibnamefont
  {Burkard}},\ }\href@noop {} {\bibfield  {journal} {\bibinfo  {journal} {Phys.
  Rev. B}\ }\textbf {\bibinfo {volume} {82}},\ \bibinfo {pages} {155424}
  (\bibinfo {year} {2010})}\BibitemShut {NoStop}%
\bibitem [{\citenamefont {Li}\ \emph {et~al.}(2014)\citenamefont {Li},
  \citenamefont {Benjamin}, \citenamefont {Briggs},\ and\ \citenamefont
  {Laird}}]{Li-edsr}%
  \BibitemOpen
  \bibfield  {author} {\bibinfo {author} {\bibfnamefont {Y.}~\bibnamefont
  {Li}}, \bibinfo {author} {\bibfnamefont {S.~C.}\ \bibnamefont {Benjamin}},
  \bibinfo {author} {\bibfnamefont {G.~A.~D.}\ \bibnamefont {Briggs}}, \ and\
  \bibinfo {author} {\bibfnamefont {E.~A.}\ \bibnamefont {Laird}},\ }\href@noop
  {} {\bibfield  {journal} {\bibinfo  {journal} {Phys. Rev. B}\ }\textbf
  {\bibinfo {volume} {90}},\ \bibinfo {pages} {195440} (\bibinfo {year}
  {2014})}\BibitemShut {NoStop}%
\bibitem [{\citenamefont {Flensberg}\ and\ \citenamefont
  {Marcus}(2010)}]{FlensbergMarcus}%
  \BibitemOpen
  \bibfield  {author} {\bibinfo {author} {\bibfnamefont {K.}~\bibnamefont
  {Flensberg}}\ and\ \bibinfo {author} {\bibfnamefont {C.~M.}\ \bibnamefont
  {Marcus}},\ }\href@noop {} {\bibfield  {journal} {\bibinfo  {journal} {Phys.
  Rev. B}\ }\textbf {\bibinfo {volume} {81}},\ \bibinfo {pages} {195418}
  (\bibinfo {year} {2010})}\BibitemShut {NoStop}%
\bibitem [{\citenamefont {Novoselov}\ \emph {et~al.}(2016)\citenamefont
  {Novoselov}, \citenamefont {Mishchenko}, \citenamefont {Carvalho},\ and\
  \citenamefont {Castro~Neto}}]{Novoselov}%
  \BibitemOpen
  \bibfield  {author} {\bibinfo {author} {\bibfnamefont {K.~S.}\ \bibnamefont
  {Novoselov}}, \bibinfo {author} {\bibfnamefont {A.}~\bibnamefont
  {Mishchenko}}, \bibinfo {author} {\bibfnamefont {A.}~\bibnamefont
  {Carvalho}}, \ and\ \bibinfo {author} {\bibfnamefont {A.~H.}\ \bibnamefont
  {Castro~Neto}},\ }\href@noop {} {\bibfield  {journal} {\bibinfo  {journal}
  {Science}\ }\textbf {\bibinfo {volume} {353}},\ \bibinfo {pages} {6298}
  (\bibinfo {year} {2016})}\BibitemShut {NoStop}%
\bibitem [{\citenamefont {Roldan}\ \emph {et~al.}(2017)\citenamefont {Roldan},
  \citenamefont {Chirolli}, \citenamefont {Prada}, \citenamefont
  {Silva-Guillen}, \citenamefont {San-Jose},\ and\ \citenamefont
  {Guinea}}]{Roldan}%
  \BibitemOpen
  \bibfield  {author} {\bibinfo {author} {\bibfnamefont {R.}~\bibnamefont
  {Roldan}}, \bibinfo {author} {\bibfnamefont {L.}~\bibnamefont {Chirolli}},
  \bibinfo {author} {\bibfnamefont {E.}~\bibnamefont {Prada}}, \bibinfo
  {author} {\bibfnamefont {J.~A.}\ \bibnamefont {Silva-Guillen}}, \bibinfo
  {author} {\bibfnamefont {P.}~\bibnamefont {San-Jose}}, \ and\ \bibinfo
  {author} {\bibfnamefont {F.}~\bibnamefont {Guinea}},\ }\href@noop {}
  {\bibfield  {journal} {\bibinfo  {journal} {Chem. Soc. Rev.}\ }\textbf
  {\bibinfo {volume} {46}},\ \bibinfo {pages} {4387} (\bibinfo {year}
  {2017})}\BibitemShut {NoStop}%
\bibitem [{\citenamefont {\ifmmode \check{Z}\else
  \v{Z}\fi{}uti\ifmmode~\acute{c}\else \'{c}\fi{}}\ \emph
  {et~al.}(2004)\citenamefont {\ifmmode \check{Z}\else
  \v{Z}\fi{}uti\ifmmode~\acute{c}\else \'{c}\fi{}}, \citenamefont {Fabian},\
  and\ \citenamefont {Das~Sarma}}]{Zutic}%
  \BibitemOpen
  \bibfield  {author} {\bibinfo {author} {\bibfnamefont {I.}~\bibnamefont
  {\ifmmode \check{Z}\else \v{Z}\fi{}uti\ifmmode~\acute{c}\else \'{c}\fi{}}},
  \bibinfo {author} {\bibfnamefont {J.}~\bibnamefont {Fabian}}, \ and\ \bibinfo
  {author} {\bibfnamefont {S.}~\bibnamefont {Das~Sarma}},\ }\href@noop {}
  {\bibfield  {journal} {\bibinfo  {journal} {Rev. Mod. Phys.}\ }\textbf
  {\bibinfo {volume} {76}},\ \bibinfo {pages} {323} (\bibinfo {year}
  {2004})}\BibitemShut {NoStop}%
\bibitem [{\citenamefont {Han}(2016)}]{Han}%
  \BibitemOpen
  \bibfield  {author} {\bibinfo {author} {\bibfnamefont {W.}~\bibnamefont
  {Han}},\ }\href@noop {} {\bibfield  {journal} {\bibinfo  {journal} {APL
  Materials}\ }\textbf {\bibinfo {volume} {4}},\ \bibinfo {pages} {032401}
  (\bibinfo {year} {2016})}\BibitemShut {NoStop}%
\bibitem [{\citenamefont {Schaibley}\ \emph {et~al.}(2016)\citenamefont
  {Schaibley}, \citenamefont {Yu}, \citenamefont {Clark}, \citenamefont
  {Rivera}, \citenamefont {Ross}, \citenamefont {Seyler}, \citenamefont {Yao},\
  and\ \citenamefont {Xu}}]{Schaibley}%
  \BibitemOpen
  \bibfield  {author} {\bibinfo {author} {\bibfnamefont {J.~R.}\ \bibnamefont
  {Schaibley}}, \bibinfo {author} {\bibfnamefont {H.}~\bibnamefont {Yu}},
  \bibinfo {author} {\bibfnamefont {G.}~\bibnamefont {Clark}}, \bibinfo
  {author} {\bibfnamefont {P.}~\bibnamefont {Rivera}}, \bibinfo {author}
  {\bibfnamefont {J.~S.}\ \bibnamefont {Ross}}, \bibinfo {author}
  {\bibfnamefont {K.~L.}\ \bibnamefont {Seyler}}, \bibinfo {author}
  {\bibfnamefont {W.}~\bibnamefont {Yao}}, \ and\ \bibinfo {author}
  {\bibfnamefont {X.}~\bibnamefont {Xu}},\ }\href@noop {} {\bibfield  {journal}
  {\bibinfo  {journal} {Nat. Rev. Materials}\ }\textbf {\bibinfo {volume}
  {1}},\ \bibinfo {pages} {16055} (\bibinfo {year} {2016})}\BibitemShut
  {NoStop}%
\bibitem [{\citenamefont {Wang}\ \emph {et~al.}(2012)\citenamefont {Wang},
  \citenamefont {Kalantar-Zadeh}, \citenamefont {Kis}, \citenamefont
  {Coleman},\ and\ \citenamefont {Strano}}]{Wang2012}%
  \BibitemOpen
  \bibfield  {author} {\bibinfo {author} {\bibfnamefont {Q.~H.}\ \bibnamefont
  {Wang}}, \bibinfo {author} {\bibfnamefont {K.}~\bibnamefont
  {Kalantar-Zadeh}}, \bibinfo {author} {\bibfnamefont {A.}~\bibnamefont {Kis}},
  \bibinfo {author} {\bibfnamefont {J.~N.}\ \bibnamefont {Coleman}}, \ and\
  \bibinfo {author} {\bibfnamefont {M.~S.}\ \bibnamefont {Strano}},\
  }\href@noop {} {\bibfield  {journal} {\bibinfo  {journal} {Nat Nano}\
  }\textbf {\bibinfo {volume} {7}},\ \bibinfo {pages} {699} (\bibinfo {year}
  {2012})}\BibitemShut {NoStop}%
\bibitem [{\citenamefont {Cappelluti}\ \emph {et~al.}(2013)\citenamefont
  {Cappelluti}, \citenamefont {Rold\'an}, \citenamefont {Silva-Guill\'en},
  \citenamefont {Ordej\'on},\ and\ \citenamefont {Guinea}}]{Cappelluti13}%
  \BibitemOpen
  \bibfield  {author} {\bibinfo {author} {\bibfnamefont {E.}~\bibnamefont
  {Cappelluti}}, \bibinfo {author} {\bibfnamefont {R.}~\bibnamefont
  {Rold\'an}}, \bibinfo {author} {\bibfnamefont {J.~A.}\ \bibnamefont
  {Silva-Guill\'en}}, \bibinfo {author} {\bibfnamefont {P.}~\bibnamefont
  {Ordej\'on}}, \ and\ \bibinfo {author} {\bibfnamefont {F.}~\bibnamefont
  {Guinea}},\ }\href@noop {} {\bibfield  {journal} {\bibinfo  {journal} {Phys.
  Rev. B}\ }\textbf {\bibinfo {volume} {88}},\ \bibinfo {pages} {075409}
  (\bibinfo {year} {2013})}\BibitemShut {NoStop}%
\bibitem [{\citenamefont {Rold{\'a}n}\ \emph {et~al.}(2014)\citenamefont
  {Rold{\'a}n}, \citenamefont {L{\'o}pez-Sancho}, \citenamefont {Guinea},
  \citenamefont {Cappelluti}, \citenamefont {Silva-Guill{\'e}n},\ and\
  \citenamefont {Ordej{\'o}n}}]{Roldan14}%
  \BibitemOpen
  \bibfield  {author} {\bibinfo {author} {\bibfnamefont {R.}~\bibnamefont
  {Rold{\'a}n}}, \bibinfo {author} {\bibfnamefont {M.~P.}\ \bibnamefont
  {L{\'o}pez-Sancho}}, \bibinfo {author} {\bibfnamefont {F.}~\bibnamefont
  {Guinea}}, \bibinfo {author} {\bibfnamefont {E.}~\bibnamefont {Cappelluti}},
  \bibinfo {author} {\bibfnamefont {J.~A.}\ \bibnamefont {Silva-Guill{\'e}n}},
  \ and\ \bibinfo {author} {\bibfnamefont {P.}~\bibnamefont {Ordej{\'o}n}},\
  }\href@noop {} {\bibfield  {journal} {\bibinfo  {journal} {2D Materials}\
  }\textbf {\bibinfo {volume} {1}},\ \bibinfo {pages} {034003} (\bibinfo {year}
  {2014})}\BibitemShut {NoStop}%
\bibitem [{\citenamefont {Korm\'anyos}\ \emph {et~al.}(2013)\citenamefont
  {Korm\'anyos}, \citenamefont {Z\'olyomi}, \citenamefont {Drummond},
  \citenamefont {Rakyta}, \citenamefont {Burkard},\ and\ \citenamefont
  {Fal'ko}}]{Kormanyos-mos2}%
  \BibitemOpen
  \bibfield  {author} {\bibinfo {author} {\bibfnamefont {A.}~\bibnamefont
  {Korm\'anyos}}, \bibinfo {author} {\bibfnamefont {V.}~\bibnamefont
  {Z\'olyomi}}, \bibinfo {author} {\bibfnamefont {N.~D.}\ \bibnamefont
  {Drummond}}, \bibinfo {author} {\bibfnamefont {P.}~\bibnamefont {Rakyta}},
  \bibinfo {author} {\bibfnamefont {G.}~\bibnamefont {Burkard}}, \ and\
  \bibinfo {author} {\bibfnamefont {V.~I.}\ \bibnamefont {Fal'ko}},\ }\href
  {\doibase 10.1103/PhysRevB.88.045416} {\bibfield  {journal} {\bibinfo
  {journal} {Phys. Rev. B}\ }\textbf {\bibinfo {volume} {88}},\ \bibinfo
  {pages} {045416} (\bibinfo {year} {2013})}\BibitemShut {NoStop}%
\bibitem [{\citenamefont {Song}\ \emph
  {et~al.}(2015{\natexlab{a}})\citenamefont {Song}, \citenamefont {Liu},
  \citenamefont {Mosallanejad}, \citenamefont {You}, \citenamefont {Han},
  \citenamefont {Chen}, \citenamefont {Li}, \citenamefont {Cao}, \citenamefont
  {Xiao}, \citenamefont {Guo},\ and\ \citenamefont {Guo}}]{Song}%
  \BibitemOpen
  \bibfield  {author} {\bibinfo {author} {\bibfnamefont {X.-X.}\ \bibnamefont
  {Song}}, \bibinfo {author} {\bibfnamefont {D.}~\bibnamefont {Liu}}, \bibinfo
  {author} {\bibfnamefont {V.}~\bibnamefont {Mosallanejad}}, \bibinfo {author}
  {\bibfnamefont {J.}~\bibnamefont {You}}, \bibinfo {author} {\bibfnamefont
  {T.-Y.}\ \bibnamefont {Han}}, \bibinfo {author} {\bibfnamefont {D.-T.}\
  \bibnamefont {Chen}}, \bibinfo {author} {\bibfnamefont {H.-O.}\ \bibnamefont
  {Li}}, \bibinfo {author} {\bibfnamefont {G.}~\bibnamefont {Cao}}, \bibinfo
  {author} {\bibfnamefont {M.}~\bibnamefont {Xiao}}, \bibinfo {author}
  {\bibfnamefont {G.-C.}\ \bibnamefont {Guo}}, \ and\ \bibinfo {author}
  {\bibfnamefont {G.-P.}\ \bibnamefont {Guo}},\ }\href@noop {} {\bibfield
  {journal} {\bibinfo  {journal} {Nanoscale}\ }\textbf {\bibinfo {volume}
  {7}},\ \bibinfo {pages} {16867} (\bibinfo {year}
  {2015}{\natexlab{a}})}\BibitemShut {NoStop}%
\bibitem [{\citenamefont {Luo}\ \emph {et~al.}(2017)\citenamefont {Luo},
  \citenamefont {Zhang}, \citenamefont {Li}, \citenamefont {Song},
  \citenamefont {Deng}, \citenamefont {Cao}, \citenamefont {Xiao},\ and\
  \citenamefont {Guo}}]{Luo2017}%
  \BibitemOpen
  \bibfield  {author} {\bibinfo {author} {\bibfnamefont {G.}~\bibnamefont
  {Luo}}, \bibinfo {author} {\bibfnamefont {Z.-Z.}\ \bibnamefont {Zhang}},
  \bibinfo {author} {\bibfnamefont {H.-O.}\ \bibnamefont {Li}}, \bibinfo
  {author} {\bibfnamefont {X.-X.}\ \bibnamefont {Song}}, \bibinfo {author}
  {\bibfnamefont {G.-W.}\ \bibnamefont {Deng}}, \bibinfo {author}
  {\bibfnamefont {G.}~\bibnamefont {Cao}}, \bibinfo {author} {\bibfnamefont
  {M.}~\bibnamefont {Xiao}}, \ and\ \bibinfo {author} {\bibfnamefont {G.-P.}\
  \bibnamefont {Guo}},\ }\href@noop {} {\bibfield  {journal} {\bibinfo
  {journal} {Frontiers of Physics}\ }\textbf {\bibinfo {volume} {12}},\
  \bibinfo {pages} {128502} (\bibinfo {year} {2017})}\BibitemShut {NoStop}%
\bibitem [{\citenamefont {Song}\ \emph
  {et~al.}(2015{\natexlab{b}})\citenamefont {Song}, \citenamefont {Zhang},
  \citenamefont {You}, \citenamefont {Liu}, \citenamefont {Li}, \citenamefont
  {Cao}, \citenamefont {Xiao},\ and\ \citenamefont {Guo}}]{Song2015}%
  \BibitemOpen
  \bibfield  {author} {\bibinfo {author} {\bibfnamefont {X.-X.}\ \bibnamefont
  {Song}}, \bibinfo {author} {\bibfnamefont {Z.-Z.}\ \bibnamefont {Zhang}},
  \bibinfo {author} {\bibfnamefont {J.}~\bibnamefont {You}}, \bibinfo {author}
  {\bibfnamefont {D.}~\bibnamefont {Liu}}, \bibinfo {author} {\bibfnamefont
  {H.-O.}\ \bibnamefont {Li}}, \bibinfo {author} {\bibfnamefont
  {G.}~\bibnamefont {Cao}}, \bibinfo {author} {\bibfnamefont {M.}~\bibnamefont
  {Xiao}}, \ and\ \bibinfo {author} {\bibfnamefont {G.-P.}\ \bibnamefont
  {Guo}},\ }\href@noop {} {\bibfield  {journal} {\bibinfo  {journal}
  {Scientific Reports}\ }\textbf {\bibinfo {volume} {5}},\ \bibinfo {pages}
  {16113 EP } (\bibinfo {year} {2015}{\natexlab{b}})}\BibitemShut {NoStop}%
\bibitem [{\citenamefont {Javaid}\ \emph {et~al.}(2017)\citenamefont {Javaid},
  \citenamefont {Drumm}, \citenamefont {Russo},\ and\ \citenamefont
  {Greentree}}]{Javaid}%
  \BibitemOpen
  \bibfield  {author} {\bibinfo {author} {\bibfnamefont {M.}~\bibnamefont
  {Javaid}}, \bibinfo {author} {\bibfnamefont {D.~W.}\ \bibnamefont {Drumm}},
  \bibinfo {author} {\bibfnamefont {S.~P.}\ \bibnamefont {Russo}}, \ and\
  \bibinfo {author} {\bibfnamefont {A.~D.}\ \bibnamefont {Greentree}},\
  }\href@noop {} {\bibfield  {journal} {\bibinfo  {journal} {Nanotechnology}\
  }\textbf {\bibinfo {volume} {28}},\ \bibinfo {pages} {125203} (\bibinfo
  {year} {2017})}\BibitemShut {NoStop}%
\bibitem [{\citenamefont {Wang}\ \emph {et~al.}()\citenamefont {Wang},
  \citenamefont {Taniguchi}, \citenamefont {Watanabe},\ and\ \citenamefont
  {Kim}}]{KeWang}%
  \BibitemOpen
  \bibfield  {author} {\bibinfo {author} {\bibfnamefont {K.}~\bibnamefont
  {Wang}}, \bibinfo {author} {\bibfnamefont {T.}~\bibnamefont {Taniguchi}},
  \bibinfo {author} {\bibfnamefont {K.}~\bibnamefont {Watanabe}}, \ and\
  \bibinfo {author} {\bibfnamefont {P.}~\bibnamefont {Kim}},\ }\href@noop {}
  {\enquote {\bibinfo {title} {Engineering quantum confinement in
  semiconducting van der waals heterostructure},}\ }\bibinfo {note}
  {ArXiv:1610.02929 (unpublished)}\BibitemShut {NoStop}%
\bibitem [{\citenamefont {Zhang}\ \emph {et~al.}(2017)\citenamefont {Zhang},
  \citenamefont {Song}, \citenamefont {Luo}, \citenamefont {Deng},
  \citenamefont {Mosallanejad}, \citenamefont {Taniguchi}, \citenamefont
  {Watanabe}, \citenamefont {Li}, \citenamefont {Cao}, \citenamefont {Guo},
  \citenamefont {Nori},\ and\ \citenamefont {Guo}}]{ZhuoZhiZhang}%
  \BibitemOpen
  \bibfield  {author} {\bibinfo {author} {\bibfnamefont {Z.-Z.}\ \bibnamefont
  {Zhang}}, \bibinfo {author} {\bibfnamefont {X.-X.}\ \bibnamefont {Song}},
  \bibinfo {author} {\bibfnamefont {G.}~\bibnamefont {Luo}}, \bibinfo {author}
  {\bibfnamefont {G.-W.}\ \bibnamefont {Deng}}, \bibinfo {author}
  {\bibfnamefont {V.}~\bibnamefont {Mosallanejad}}, \bibinfo {author}
  {\bibfnamefont {T.}~\bibnamefont {Taniguchi}}, \bibinfo {author}
  {\bibfnamefont {K.}~\bibnamefont {Watanabe}}, \bibinfo {author}
  {\bibfnamefont {H.-O.}\ \bibnamefont {Li}}, \bibinfo {author} {\bibfnamefont
  {G.}~\bibnamefont {Cao}}, \bibinfo {author} {\bibfnamefont {G.-C.}\
  \bibnamefont {Guo}}, \bibinfo {author} {\bibfnamefont {F.}~\bibnamefont
  {Nori}}, \ and\ \bibinfo {author} {\bibfnamefont {G.-P.}\ \bibnamefont
  {Guo}},\ }\href@noop {} {\bibfield  {journal} {\bibinfo  {journal} {Sci.
  Adv.}\ }\textbf {\bibinfo {volume} {3}},\ \bibinfo {pages} {1701699}
  (\bibinfo {year} {2017})}\BibitemShut {NoStop}%
\bibitem [{\citenamefont {Klinovaja}\ and\ \citenamefont
  {Loss}(2013)}]{Klinovaja}%
  \BibitemOpen
  \bibfield  {author} {\bibinfo {author} {\bibfnamefont {J.}~\bibnamefont
  {Klinovaja}}\ and\ \bibinfo {author} {\bibfnamefont {D.}~\bibnamefont
  {Loss}},\ }\href {\doibase 10.1103/PhysRevB.88.075404} {\bibfield  {journal}
  {\bibinfo  {journal} {Phys. Rev. B}\ }\textbf {\bibinfo {volume} {88}},\
  \bibinfo {pages} {075404} (\bibinfo {year} {2013})}\BibitemShut {NoStop}%
\bibitem [{\citenamefont {Korm\'anyos}\ \emph {et~al.}(2014)\citenamefont
  {Korm\'anyos}, \citenamefont {Z\'olyomi}, \citenamefont {Drummond},\ and\
  \citenamefont {Burkard}}]{Kormanyos-mos2quantumdots}%
  \BibitemOpen
  \bibfield  {author} {\bibinfo {author} {\bibfnamefont {A.}~\bibnamefont
  {Korm\'anyos}}, \bibinfo {author} {\bibfnamefont {V.}~\bibnamefont
  {Z\'olyomi}}, \bibinfo {author} {\bibfnamefont {N.~D.}\ \bibnamefont
  {Drummond}}, \ and\ \bibinfo {author} {\bibfnamefont {G.}~\bibnamefont
  {Burkard}},\ }\href {\doibase 10.1103/PhysRevX.4.011034} {\bibfield
  {journal} {\bibinfo  {journal} {Phys. Rev. X}\ }\textbf {\bibinfo {volume}
  {4}},\ \bibinfo {pages} {011034} (\bibinfo {year} {2014})}\BibitemShut
  {NoStop}%
\bibitem [{\citenamefont {Liu}\ \emph {et~al.}(2014)\citenamefont {Liu},
  \citenamefont {Pang}, \citenamefont {Yao},\ and\ \citenamefont
  {Yao}}]{GuiBinLiu_intervalley}%
  \BibitemOpen
  \bibfield  {author} {\bibinfo {author} {\bibfnamefont {G.-B.}\ \bibnamefont
  {Liu}}, \bibinfo {author} {\bibfnamefont {H.}~\bibnamefont {Pang}}, \bibinfo
  {author} {\bibfnamefont {Y.}~\bibnamefont {Yao}}, \ and\ \bibinfo {author}
  {\bibfnamefont {W.}~\bibnamefont {Yao}},\ }\href
  {http://stacks.iop.org/1367-2630/16/i=10/a=105011} {\bibfield  {journal}
  {\bibinfo  {journal} {New Journal of Physics}\ }\textbf {\bibinfo {volume}
  {16}},\ \bibinfo {pages} {105011} (\bibinfo {year} {2014})}\BibitemShut
  {NoStop}%
\bibitem [{\citenamefont {Wu}\ \emph {et~al.}(2016)\citenamefont {Wu},
  \citenamefont {Tong}, \citenamefont {Liu}, \citenamefont {Yu},\ and\
  \citenamefont {Yao}}]{YueWu_spinvalley}%
  \BibitemOpen
  \bibfield  {author} {\bibinfo {author} {\bibfnamefont {Y.}~\bibnamefont
  {Wu}}, \bibinfo {author} {\bibfnamefont {Q.}~\bibnamefont {Tong}}, \bibinfo
  {author} {\bibfnamefont {G.-B.}\ \bibnamefont {Liu}}, \bibinfo {author}
  {\bibfnamefont {H.}~\bibnamefont {Yu}}, \ and\ \bibinfo {author}
  {\bibfnamefont {W.}~\bibnamefont {Yao}},\ }\href {\doibase
  10.1103/PhysRevB.93.045313} {\bibfield  {journal} {\bibinfo  {journal} {Phys.
  Rev. B}\ }\textbf {\bibinfo {volume} {93}},\ \bibinfo {pages} {045313}
  (\bibinfo {year} {2016})}\BibitemShut {NoStop}%
\bibitem [{\citenamefont {Pearce}\ and\ \citenamefont
  {Burkard}(2017)}]{Pearce_Burkard}%
  \BibitemOpen
  \bibfield  {author} {\bibinfo {author} {\bibfnamefont {A.~J.}\ \bibnamefont
  {Pearce}}\ and\ \bibinfo {author} {\bibfnamefont {G.}~\bibnamefont
  {Burkard}},\ }\href {http://stacks.iop.org/2053-1583/4/i=2/a=025114}
  {\bibfield  {journal} {\bibinfo  {journal} {2D Materials}\ }\textbf {\bibinfo
  {volume} {4}},\ \bibinfo {pages} {025114} (\bibinfo {year}
  {2017})}\BibitemShut {NoStop}%
\bibitem [{\citenamefont {Brooks}\ and\ \citenamefont
  {Burkard}(2017)}]{BrooksBurkard}%
  \BibitemOpen
  \bibfield  {author} {\bibinfo {author} {\bibfnamefont {M.}~\bibnamefont
  {Brooks}}\ and\ \bibinfo {author} {\bibfnamefont {G.}~\bibnamefont
  {Burkard}},\ }\href {\doibase 10.1103/PhysRevB.95.245411} {\bibfield
  {journal} {\bibinfo  {journal} {Phys. Rev. B}\ }\textbf {\bibinfo {volume}
  {95}},\ \bibinfo {pages} {245411} (\bibinfo {year} {2017})}\BibitemShut
  {NoStop}%
\bibitem [{\citenamefont {Itoh}\ and\ \citenamefont
  {Watanabe}(2014)}]{Itoh_isotope}%
  \BibitemOpen
  \bibfield  {author} {\bibinfo {author} {\bibfnamefont {K.~M.}\ \bibnamefont
  {Itoh}}\ and\ \bibinfo {author} {\bibfnamefont {H.}~\bibnamefont
  {Watanabe}},\ }\href {\doibase 10.1557/mrc.2014.32} {\bibfield  {journal}
  {\bibinfo  {journal} {MRS Communications}\ }\textbf {\bibinfo {volume} {4}},\
  \bibinfo {pages} {143} (\bibinfo {year} {2014})}\BibitemShut {NoStop}%
\bibitem [{\citenamefont {{Kaasbjerg}}\ \emph {et~al.}(2016)\citenamefont
  {{Kaasbjerg}}, \citenamefont {{Low}},\ and\ \citenamefont
  {{Jauho}}}]{Kaasbjerg2016}%
  \BibitemOpen
  \bibfield  {author} {\bibinfo {author} {\bibfnamefont {K.}~\bibnamefont
  {{Kaasbjerg}}}, \bibinfo {author} {\bibfnamefont {T.}~\bibnamefont {{Low}}},
  \ and\ \bibinfo {author} {\bibfnamefont {A.-P.}\ \bibnamefont {{Jauho}}},\
  }\href@noop {} {\bibfield  {journal} {\bibinfo  {journal} {ArXiv e-prints}\ }
  (\bibinfo {year} {2016})},\ \Eprint {http://arxiv.org/abs/1612.00469}
  {arXiv:1612.00469 [cond-mat.mes-hall]} \BibitemShut {NoStop}%
\bibitem [{\citenamefont {Kaasbjerg}\ \emph {et~al.}(2017)\citenamefont
  {Kaasbjerg}, \citenamefont {Martiny}, \citenamefont {Low},\ and\
  \citenamefont {Jauho}}]{Kaasbjerg2017}%
  \BibitemOpen
  \bibfield  {author} {\bibinfo {author} {\bibfnamefont {K.}~\bibnamefont
  {Kaasbjerg}}, \bibinfo {author} {\bibfnamefont {J.~H.~J.}\ \bibnamefont
  {Martiny}}, \bibinfo {author} {\bibfnamefont {T.}~\bibnamefont {Low}}, \ and\
  \bibinfo {author} {\bibfnamefont {A.-P.}\ \bibnamefont {Jauho}},\ }\href@noop
  {} {\bibfield  {journal} {\bibinfo  {journal} {Phys. Rev. B}\ }\textbf
  {\bibinfo {volume} {96}},\ \bibinfo {pages} {241411} (\bibinfo {year}
  {2017})}\BibitemShut {NoStop}%
\bibitem [{\citenamefont {Hong}\ \emph {et~al.}(2015)\citenamefont {Hong},
  \citenamefont {Hu}, \citenamefont {Probert}, \citenamefont {Li},
  \citenamefont {Lv}, \citenamefont {Yang}, \citenamefont {Gu}, \citenamefont
  {Mao}, \citenamefont {Feng}, \citenamefont {Xie}, \citenamefont {Zhang},
  \citenamefont {Wu}, \citenamefont {Zhang}, \citenamefont {Jin}, \citenamefont
  {Ji}, \citenamefont {Zhang}, \citenamefont {Yuan},\ and\ \citenamefont
  {Zhang}}]{JinhuaHong}%
  \BibitemOpen
  \bibfield  {author} {\bibinfo {author} {\bibfnamefont {J.}~\bibnamefont
  {Hong}}, \bibinfo {author} {\bibfnamefont {Z.}~\bibnamefont {Hu}}, \bibinfo
  {author} {\bibfnamefont {M.}~\bibnamefont {Probert}}, \bibinfo {author}
  {\bibfnamefont {K.}~\bibnamefont {Li}}, \bibinfo {author} {\bibfnamefont
  {D.}~\bibnamefont {Lv}}, \bibinfo {author} {\bibfnamefont {X.}~\bibnamefont
  {Yang}}, \bibinfo {author} {\bibfnamefont {L.}~\bibnamefont {Gu}}, \bibinfo
  {author} {\bibfnamefont {N.}~\bibnamefont {Mao}}, \bibinfo {author}
  {\bibfnamefont {Q.}~\bibnamefont {Feng}}, \bibinfo {author} {\bibfnamefont
  {L.}~\bibnamefont {Xie}}, \bibinfo {author} {\bibfnamefont {J.}~\bibnamefont
  {Zhang}}, \bibinfo {author} {\bibfnamefont {D.}~\bibnamefont {Wu}}, \bibinfo
  {author} {\bibfnamefont {Z.}~\bibnamefont {Zhang}}, \bibinfo {author}
  {\bibfnamefont {C.}~\bibnamefont {Jin}}, \bibinfo {author} {\bibfnamefont
  {W.}~\bibnamefont {Ji}}, \bibinfo {author} {\bibfnamefont {X.}~\bibnamefont
  {Zhang}}, \bibinfo {author} {\bibfnamefont {J.}~\bibnamefont {Yuan}}, \ and\
  \bibinfo {author} {\bibfnamefont {Z.}~\bibnamefont {Zhang}},\ }\href@noop {}
  {\bibfield  {journal} {\bibinfo  {journal} {Nat Commun}\ }\textbf {\bibinfo
  {volume} {6}},\ \bibinfo {pages} {6293} (\bibinfo {year} {2015})}\BibitemShut
  {NoStop}%
\bibitem [{\citenamefont {Zhou}\ \emph {et~al.}(2013)\citenamefont {Zhou},
  \citenamefont {Zou}, \citenamefont {Najmaei}, \citenamefont {Liu},
  \citenamefont {Shi}, \citenamefont {Kong}, \citenamefont {Lou}, \citenamefont
  {Ajayan}, \citenamefont {Yakobson},\ and\ \citenamefont {Idrobo}}]{zhou}%
  \BibitemOpen
  \bibfield  {author} {\bibinfo {author} {\bibfnamefont {W.}~\bibnamefont
  {Zhou}}, \bibinfo {author} {\bibfnamefont {X.}~\bibnamefont {Zou}}, \bibinfo
  {author} {\bibfnamefont {S.}~\bibnamefont {Najmaei}}, \bibinfo {author}
  {\bibfnamefont {Z.}~\bibnamefont {Liu}}, \bibinfo {author} {\bibfnamefont
  {Y.}~\bibnamefont {Shi}}, \bibinfo {author} {\bibfnamefont {J.}~\bibnamefont
  {Kong}}, \bibinfo {author} {\bibfnamefont {J.}~\bibnamefont {Lou}}, \bibinfo
  {author} {\bibfnamefont {P.~M.}\ \bibnamefont {Ajayan}}, \bibinfo {author}
  {\bibfnamefont {B.~I.}\ \bibnamefont {Yakobson}}, \ and\ \bibinfo {author}
  {\bibfnamefont {J.-C.}\ \bibnamefont {Idrobo}},\ }\href {\doibase
  10.1021/nl4007479} {\bibfield  {journal} {\bibinfo  {journal} {Nano Letters}\
  }\textbf {\bibinfo {volume} {13}},\ \bibinfo {pages} {2615} (\bibinfo {year}
  {2013})}\BibitemShut {NoStop}%
\bibitem [{\citenamefont {Lin}\ \emph {et~al.}(2015)\citenamefont {Lin},
  \citenamefont {Bj{\"o}rkman}, \citenamefont {Komsa}, \citenamefont {Teng},
  \citenamefont {Yeh}, \citenamefont {Huang}, \citenamefont {Lin},
  \citenamefont {Jadczak}, \citenamefont {Huang}, \citenamefont {Chiu},
  \citenamefont {Krasheninnikov},\ and\ \citenamefont {Suenaga}}]{Lin2015}%
  \BibitemOpen
  \bibfield  {author} {\bibinfo {author} {\bibfnamefont {Y.-C.}\ \bibnamefont
  {Lin}}, \bibinfo {author} {\bibfnamefont {T.}~\bibnamefont {Bj{\"o}rkman}},
  \bibinfo {author} {\bibfnamefont {H.-P.}\ \bibnamefont {Komsa}}, \bibinfo
  {author} {\bibfnamefont {P.-Y.}\ \bibnamefont {Teng}}, \bibinfo {author}
  {\bibfnamefont {C.-H.}\ \bibnamefont {Yeh}}, \bibinfo {author} {\bibfnamefont
  {F.-S.}\ \bibnamefont {Huang}}, \bibinfo {author} {\bibfnamefont {K.-H.}\
  \bibnamefont {Lin}}, \bibinfo {author} {\bibfnamefont {J.}~\bibnamefont
  {Jadczak}}, \bibinfo {author} {\bibfnamefont {Y.-S.}\ \bibnamefont {Huang}},
  \bibinfo {author} {\bibfnamefont {P.-W.}\ \bibnamefont {Chiu}}, \bibinfo
  {author} {\bibfnamefont {A.~V.}\ \bibnamefont {Krasheninnikov}}, \ and\
  \bibinfo {author} {\bibfnamefont {K.}~\bibnamefont {Suenaga}},\ }\href@noop
  {} {\bibfield  {journal} {\bibinfo  {journal} {Nature Communications}\
  }\textbf {\bibinfo {volume} {6}},\ \bibinfo {pages} {6736} (\bibinfo {year}
  {2015})}\BibitemShut {NoStop}%
\bibitem [{\citenamefont {Vancs{\'o}}\ \emph {et~al.}(2016)\citenamefont
  {Vancs{\'o}}, \citenamefont {Magda}, \citenamefont {Pet{\H o}}, \citenamefont
  {Noh}, \citenamefont {Kim}, \citenamefont {Hwang}, \citenamefont {Bir{\'o}},\
  and\ \citenamefont {Tapaszt{\'o}}}]{Vancso}%
  \BibitemOpen
  \bibfield  {author} {\bibinfo {author} {\bibfnamefont {P.}~\bibnamefont
  {Vancs{\'o}}}, \bibinfo {author} {\bibfnamefont {G.~Z.}\ \bibnamefont
  {Magda}}, \bibinfo {author} {\bibfnamefont {J.}~\bibnamefont {Pet{\H o}}},
  \bibinfo {author} {\bibfnamefont {J.-Y.}\ \bibnamefont {Noh}}, \bibinfo
  {author} {\bibfnamefont {Y.-S.}\ \bibnamefont {Kim}}, \bibinfo {author}
  {\bibfnamefont {C.}~\bibnamefont {Hwang}}, \bibinfo {author} {\bibfnamefont
  {L.}~\bibnamefont {Bir{\'o}}}, \ and\ \bibinfo {author} {\bibfnamefont
  {L.}~\bibnamefont {Tapaszt{\'o}}},\ }\href@noop {} {\bibfield  {journal}
  {\bibinfo  {journal} {Sci. Rep.}\ }\textbf {\bibinfo {volume} {6}},\ \bibinfo
  {pages} {29726} (\bibinfo {year} {2016})}\BibitemShut {NoStop}%
\bibitem [{\citenamefont {P\'alyi}\ and\ \citenamefont
  {Burkard}(2009)}]{Palyi_hyperfine}%
  \BibitemOpen
  \bibfield  {author} {\bibinfo {author} {\bibfnamefont {A.}~\bibnamefont
  {P\'alyi}}\ and\ \bibinfo {author} {\bibfnamefont {G.}~\bibnamefont
  {Burkard}},\ }\href {\doibase 10.1103/PhysRevB.80.201404} {\bibfield
  {journal} {\bibinfo  {journal} {Phys. Rev. B}\ }\textbf {\bibinfo {volume}
  {80}},\ \bibinfo {pages} {201404} (\bibinfo {year} {2009})}\BibitemShut
  {NoStop}%
\bibitem [{\citenamefont {Sz\'echenyi}\ and\ \citenamefont
  {P\'alyi}(2014)}]{Szechenyi-maximalrabi}%
  \BibitemOpen
  \bibfield  {author} {\bibinfo {author} {\bibfnamefont {G.}~\bibnamefont
  {Sz\'echenyi}}\ and\ \bibinfo {author} {\bibfnamefont {A.}~\bibnamefont
  {P\'alyi}},\ }\href {\doibase 10.1103/PhysRevB.89.115409} {\bibfield
  {journal} {\bibinfo  {journal} {Phys. Rev. B}\ }\textbf {\bibinfo {volume}
  {89}},\ \bibinfo {pages} {115409} (\bibinfo {year} {2014})}\BibitemShut
  {NoStop}%
\bibitem [{\citenamefont {Winkler}(2003)}]{Winkler}%
  \BibitemOpen
  \bibfield  {author} {\bibinfo {author} {\bibfnamefont {R.}~\bibnamefont
  {Winkler}},\ }\href@noop {} {\emph {\bibinfo {title} {Spin-Orbit Coupling
  Effects in Two-Dimensional Electron and Hole Systems}}}\ (\bibinfo
  {publisher} {Springer-Verlag},\ \bibinfo {year} {2003})\BibitemShut {NoStop}%
\bibitem [{\citenamefont {Romh\'anyi}\ \emph {et~al.}(2015)\citenamefont
  {Romh\'anyi}, \citenamefont {Burkard},\ and\ \citenamefont
  {P\'alyi}}]{Romhanyi}%
  \BibitemOpen
  \bibfield  {author} {\bibinfo {author} {\bibfnamefont {J.}~\bibnamefont
  {Romh\'anyi}}, \bibinfo {author} {\bibfnamefont {G.}~\bibnamefont {Burkard}},
  \ and\ \bibinfo {author} {\bibfnamefont {A.}~\bibnamefont {P\'alyi}},\ }\href
  {\doibase 10.1103/PhysRevB.92.054422} {\bibfield  {journal} {\bibinfo
  {journal} {Phys. Rev. B}\ }\textbf {\bibinfo {volume} {92}},\ \bibinfo
  {pages} {054422} (\bibinfo {year} {2015})}\BibitemShut {NoStop}%
\bibitem [{\citenamefont {Crippa}\ \emph {et~al.}()\citenamefont {Crippa},
  \citenamefont {Maurand}, \citenamefont {Bourdet}, \citenamefont
  {Kotekar-Patil}, \citenamefont {Amisse}, \citenamefont {Jehl}, \citenamefont
  {Sanquer}, \citenamefont {Lavi{\'e}ville}, \citenamefont {Bohuslavskyi},
  \citenamefont {Hutin}, \citenamefont {Barraud}, \citenamefont {Vinet},
  \citenamefont {Niquet},\ and\ \citenamefont {Franceschi}}]{Crippa}%
  \BibitemOpen
  \bibfield  {author} {\bibinfo {author} {\bibfnamefont {A.}~\bibnamefont
  {Crippa}}, \bibinfo {author} {\bibfnamefont {R.}~\bibnamefont {Maurand}},
  \bibinfo {author} {\bibfnamefont {L.}~\bibnamefont {Bourdet}}, \bibinfo
  {author} {\bibfnamefont {D.}~\bibnamefont {Kotekar-Patil}}, \bibinfo {author}
  {\bibfnamefont {A.}~\bibnamefont {Amisse}}, \bibinfo {author} {\bibfnamefont
  {X.}~\bibnamefont {Jehl}}, \bibinfo {author} {\bibfnamefont {M.}~\bibnamefont
  {Sanquer}}, \bibinfo {author} {\bibfnamefont {R.}~\bibnamefont
  {Lavi{\'e}ville}}, \bibinfo {author} {\bibfnamefont {H.}~\bibnamefont
  {Bohuslavskyi}}, \bibinfo {author} {\bibfnamefont {L.}~\bibnamefont {Hutin}},
  \bibinfo {author} {\bibfnamefont {S.}~\bibnamefont {Barraud}}, \bibinfo
  {author} {\bibfnamefont {M.}~\bibnamefont {Vinet}}, \bibinfo {author}
  {\bibfnamefont {Y.-M.}\ \bibnamefont {Niquet}}, \ and\ \bibinfo {author}
  {\bibfnamefont {S.~D.}\ \bibnamefont {Franceschi}},\ }\href@noop {} {\enquote
  {\bibinfo {title} {Electrical spin driving by g-matrix modulation in
  spin-orbit qubits},}\ }\bibinfo {note} {ArXiv:1710.08690
  (unpublished)}\BibitemShut {NoStop}%
\end{thebibliography}%

\end{document}